# Convection in Binary Fluid Mixtures.
# I. Extended Traveling Wave and Stationary States.

W. Barten* and M. Lücke

*Institut für Theoretische Physik, Universität des Saarlandes, D-66041 Saarbrücken, Germany*

M. Kamps

*HLRZ, Forschungszentrum, D-52425 Jülich, Germany*

R. Schmitz

*Institut für Festkörperforschung, Forschungszentrum, D-52425 Jülich, Germany*

Nonlinear, spatially extended structures of convection rolls in horizontal layers of binary fluids heated from below are investigated in quantitative detail as a function of Rayleigh number for several negative and positive Soret coupling strengths (separation ratios) and different Lewis and Prandtl numbers characterizing different mixtures. A finite difference method was used to solve the full hydrodynamic field equations in a vertical cross section perpendicular to the roll axes, subject to realistic horizontal and laterally periodic boundary conditions in a range of experimentally accessible parameters. We elucidate the important role that the concentration field plays in the structural dynamics of the nonlinear states of stationary overturning convection (SOC) and of traveling wave (TW) convection investigated here. Structural differences in the concentration boundary layers and of the concentration plumes in TW's and SOC's and their physical consequences are discussed. These properties show that the states considered here are indeed strongly nonlinear, as expected from the magnitude of advection and diffusion in the concentration balance. The bifurcation behaviour of the states is analysed using different order parameters such as flow intensity, Nusselt number, a newly defined mixing parameter characterized by the variance of the concentration field, and the TW frequency. For further comparison with experiments, light intensity distributions are determined that can be observed in side-view shadowgraphs done with horizontal light along the roll axes. Furthermore, detailed structural analyses of all fields are made using colour coded isoplots, vertical and lateral field profiles, and lateral Fourier decompositions. They reveal, among other things, that the mirror-glide operation of lateral translation by half a wavelength combined with vertical reflection at the horizontal midplane of the layer is the longest persistent symmetry of TW and SOC states. Transport properties of the TW state are also discussed, in particular the mean lateral concentration current that is caused by the phase difference between concentration wave and velocity wave and that is roughly proportional to the TW frequency. This current plays an important role in the structural dynamics and stability of the spatially-localized traveling-wave convection investigated in an accompanying paper.

47.20.-k,47.54.+r,47.15.-x,47.10.+g

---

*Present address: Paul Scherrer Institut, CH-5232 Villigen PSI, Switzerland

# I. INTRODUCTION

Thermally-driven convection in a binary fluid of two miscible components like ethanol-water or $^3$He-$^4$He has attracted much research activity lately. It has all the qualities of a paradigmatic system for studying problems related to hydrodynamic instabilities, bifurcation theory, structure formation, complex spatio-temporal behavior, and turbulence [1]: The experimental realization is sufficiently simple to allow for controlled experiments, and the governing field equations are well known. Furthermore, experiments [2–45] and theoretical approaches [46–77] that capture properties of the experimental setups, albeit to quite a varying degree, have observed an enormous richness of spatio-temporal phenomena. Here we investigate, by numerical integration of the hydrodynamic field equations, spatially-periodic, extended convective structures consisting of straight rolls. They appear either as stationary patterns that we call states of stationary overturning convection (SOC) rolls for historical reasons [2] or as traveling-wave (TW) patterns of propagating rolls. The analysis of these extended states is a useful and indeed necessary basis for an understanding of more complex structures like localized traveling wave (LTW) states, which we investigate in an accompanying paper [46].

In particular, we want to elucidate the role of the concentration field. It is the combination (i) of the linear coupling of its degrees of freedom to the temperature field via the Soret effect and (ii) of its contribution to the buoyancy force — the concentration field directly influences the driving mechanism for convection — that causes the additional richness of convective dynamics and structures in mixtures in comparison to pure one-component fluids. The latter exhibit as a first instability just a *forwards* bifurcation of the quiescent conductive state into stationary rolls. On the one hand, concentration is advected, and since the concentration diffusion constant is small it is transported almost passively by the flow, except for boundary layer effects. But on the other hand the structural dynamics of the concentration field actively feeds back via the associated density variations into the buoyancy force that drives convection. The Soret coupling constant $\psi$ can be positive or negative depending on mean temperature and concentration, and for room temperature ethanol-water mixtures $\psi$-values between about $-0.5$ and $+0.2$ can be easily realized experimentally [3]. Without Soret coupling, $\psi = 0$, concentration deviations from the mean eventually diffuse away. For $\psi \neq 0$, however, the externally-imposed temperature stress sustains concentration variations against the mixing action of convection via the Soret effect; these in turn alter the driving force for convection. For negative Soret coupling, $\psi < 0$, this interplay causes the appearance of a Hopf bifurcation into a pair of symmetry degenerated left and right TW states in addition to the bifurcation of SOC rolls. Both of these convective solution branches typically show a subcritical bifurcation topology — the bifurcation is forwards only in the immediate vicinity of the pure-fluid limit of vanishing Soret coupling, $\psi = 0$. The associated discontinuous onset of convection reflects the nonlinear feed back mechanism: convective homogenization of the concentration distribution reduces the adverse buoyancy-depressing Soret-induced conductive concentration gradient and thus amplifies buoyancy and convection intensity.

We have calculated and analysed various states for negative as well as for positive Soret coupling parameters. However, our main interest has been focussed on the experimentally observable TW and SOC states that are located at $\psi < 0$ on the upper solution branch of the bifurcation curves of, say, flow amplitude versus Rayleigh number. We should like to stress that these experimentally observable convective states are strongly nonlinear. Their spatio-temporal behaviour can neither be described quantitatively by a weakly nonlinear expansion around their bifurcation thresholds nor by simple heuristic amplitude equation approaches. We have therefore solved the full hydrodynamic field equations numerically with realistic boundary conditions using an explicit finite-difference method. We restricted ourselves to 2d convection in the form of straight rolls by suppressing spatial variations along the roll axes. In most cases, we enforced the wavelength of the states by imposing laterally periodic boundary conditions with the cell length equal to twice the height of the fluid layer, i.e., close to the wavelength seen in experiments.

Since a review [47] of the early literature on convection in binary mixtures and a critical appraisal [1] of many of the more recent works is available – see also [48,49] – we add here only some short complementary comments that are mostly related to experimental and theoretical developments relevant to our work. After the first unambiguous visual observation [4] of TW's in an ethanol-water mixture, the use of annular convection channels [5,6] that (i) eliminated endwall effects and (ii) enforced lateral periodicity facilitated comparisons with theoretical and numerical approaches. However, on the theoretical side, amplitude-equation models and few-mode Galerkin approximations (even after strongly increasing the number of modes [7]) have failed to properly describe nonlinear TW's, e.g., their frequency variation with Rayleigh number, their bifurcation behaviour on the upper stable branch, and their field structure. So far, only Bensimon et al. [50–52] succeeded in calculating TW's on the upper branch with a saddle connection to the lower backwards bifurcating branch using a boundary layer model [78] for the concentration field. But this approach also poses some problems.

On the numerical side, simulations of so called thermosolutal convection by Deane et al. [53] yielded TW and SOC solutions. However, they are not fully applicable for quantitative comparison with the experiments [2–45] in binary mixtures for several reasons [1]. Stable nonlinear TW's comparable to the experimental ones [5,6] have been obtained numerically in Ref. [54] using a finite difference method. This analysis [54] showed the existence of large-scale mean Eulerian lateral currents of heat and concentration and a nonzero mean velocity field that had previously been found in few-mode Galerkin models by Linz et al. [55]. Also, the Lagrangian motion in passive tracer transport [8,9] in TW states is now well understood [55,56,79,57,58,10] and is different from the above mentioned Eulerian currents and fields. Using a finite-difference method at $\psi = -0.11$, Yahata [59] calculated an SOC state on the upper solution branch showing boundary layer behaviour in the concentration field. A comparison with a 52-mode Galerkin model revealed that temperature and velocity fields could very well be described by few modes, but that the concentration field was not reproduced properly by this mode ansatz. Later, Yahata [60] calculated oscillatory convection that was spatially confined to the vicinity of a lateral side wall. He also presented the concentration field, which had a similar structure as in an infinite system.

Analysing experiments in an annular channel at $\psi = -0.25$, Ohlsen et al. [11] found good agreement with the concentration boundary layer model of Bensimon et al. [51] and with numerical simulations [57] of the TW states on the upper branch all the way between the saddle and the nonhysteretic transition TW $\to$ SOC. In short, straight rectangular channels, the lateral boundaries make this transition hysteretic with a jump in TW frequency [12,13]. A comparison of experimental top-view shadowgraph images of TW and SOC states [14] with numerical results [57] also allowed the identification of the contribution from the concentration field to the light intensity profile caused by variations in the refractive index of the mixture. This seems to have been the first time that structural properties of the concentration field – albeit vertically averaged – in binary fluid convection have been observed experimentally. Subsequently, Winkler and Kolodner [15] extracted more direct information on the lateral structure of the concentration distribution in TW and SOC states from side-view shadowgraph images obtaining good agreement with numerical results [57]. On the other hand, the shadowgraph measurements of Liu and de Bruyn [16], done with very narrow (0.3$d$ in $y$-direction) and short rectangular channels, showed significant quantitative differences in intensities and bifurcation behavior with experiments in broader annular channels [14,15] and with our 2d numerical calculations. The differences can be traced back in large part to the cell geometry. In fact, our calculations agree well with experiments in channels of moderate width (1.3 to 3 times the height) [15,14]. In very broad channels, straight rolls undergo 3d instabilities [80,61,17,18]. In very narrow ones, the influence of the walls becomes too strong [16]; in addition, the critical temperature difference between the horizontal plates increases so strongly that non Oberbeck-Boussinesq effects become important. We should mention also that Zimmermann et al. [19–21] did side-view differential interferometry in ethanol-water mixtures. They found that, for mean fluid temperatures of about 10$^o$C, the existence range of stable TW's is strongly reduced and at a mean temperature of about 5$^o$C, there were no stable TW's.

Extended convective states also have been investigated for positive Soret coupling $\psi$. For a survey, see, e.g., [22,23,62]. Lhost and Platten [24] observed, in a *narrow* rectangular cell close to onset, very big rolls, as one would expect from a linear stability analysis which yields a vanishing critical wavenumber. Various experiments [25,22,26,13,23] done at not-too-small positive $\psi$ in *broad* rectangular and cylindrical cells found stationary, square-shaped convective structures. In agreement with the amplitude-equation model of Clune and Knobloch [62], they occur in the so-called Soret regime between the onset $r_{stat}$ and the reduced Rayleigh number $r = 1$ at which SOC rolls bifurcate in pure fluids. However, the channel walls of *narrow* straight or annular cells suppress square patterns and favour SOC rolls with axes oriented perpendicular to the walls. In the so-called Rayleigh region above $r \approx 1$, square convection loses stability to a stable oscillatory state [26,63,23]. Müller and Lücke [63] obtained such temporal oscillations between rolls of perpendicular orientation, using a 10-mode Lorenz model with a phenomenological sidewall forcing. With further increase of the Rayleigh number, convection in the mixture approaches the behavior in a pure fluid, i.e., SOC rolls of one direction are stabilized. To study square convection and the above described oscillatory state, Bestehorn [64] has numerically integrated the 3d field equations in the limit of infinite Prandtl number, albeit with very moderate spatial resolution of the concentration boundary layers.

Our work is organized as follows: Section II describes the system we study, the field equations, the role of the Soret coupling, and the bifurcation thresholds of convective solutions from the quiescent conductive state. In section III, we analyse, for negative Soret coupling $\psi$, the bifurcation behaviour of nonlinear TW and SOC states and their field structure and symmetry properties, especially focussing on the concentration field and the transport properties. In section IV, we discuss how structure and bifurcation behaviour change with $\psi$, and section V presents stationary states for positive Soret coupling. In section VI, we investigate variations with changing Lewis and Prandtl numbers. The conclusion in section VII lists our main results on nonlinear extended SOC and TW states. In the accompanying paper [46], we investigate spatially-localized traveling-wave convection.

## II. THE SYSTEM

We consider a horizontal layer of a binary fluid mixture like alcohol-water or normal $^3$He-$^4$He under a homogeneous gravitational field, $\mathbf{g} = -g\,\mathbf{e}_z$, that is directed downwards. A positive temperature difference,

$\Delta \underline{T} = \underline{T}_{lower} - \underline{T}_{upper}$, between the lower and upper confining boundaries is imposed externally, e.g., via highly-conducting plates in experiments. When the resulting buoyancy force (cf. below) exceeds a threshold, convective motion sets in. We consider here convection in the form of straight parallel rolls as seen in many experiments. Ignoring variations along the roll axes, we describe 2d convection in an $x - z$ plane perpendicular to the axes.

### A. Equations

The system is described by the balance equations for mass, heat, concentration, and momentum in Oberbeck-Boussinesq approximation [81–83,47]

$$0 = -\boldsymbol{\nabla} \cdot \mathbf{u}, \qquad (2.1)$$

$$\partial_t \, \delta T = -\boldsymbol{\nabla} \cdot \mathbf{Q} \; ; \; \mathbf{Q} = \mathbf{u}\, \delta T - \boldsymbol{\nabla} \, \delta T, \qquad (2.2)$$

$$\partial_t \, \delta C = -\boldsymbol{\nabla} \cdot \mathbf{J} \; ;$$
$$\mathbf{J} = \mathbf{u}\, \delta C - L \boldsymbol{\nabla} \left( \delta C - \psi \, \delta T \right), \qquad (2.3)$$

$$\partial_t \, \mathbf{u} = -\boldsymbol{\nabla}\left(\mathbf{u} : \mathbf{u} + p - \sigma \boldsymbol{\nabla} : \mathbf{u}\right) + \mathbf{B} \; ;$$
$$\mathbf{B} = \sigma\, R\, (\delta T + \delta C)\mathbf{e}_z. \qquad (2.4)$$

Lengths are scaled with the height $d$ of the layer, the time with the vertical thermal diffusion time $d^2/\kappa$, and the velocity field $\mathbf{u} = (u, 0, w)$ with $\kappa/d$. Here, $\kappa$ is the thermal diffusivity of the mixture. $\delta T = (\underline{T} - \underline{T}_0)/\Delta \underline{T}$ denotes the deviation of the temperature from the mean temperature $\underline{T}_0$ in the fluid scaled by the temperature difference between the plates. Some of the unscaled quantities are underlined for distinction from the scaled ones. The field $\delta C = (\underline{C} - \underline{C}_0)\, \beta\, /\, (\alpha\, \Delta \underline{T})$ is the scaled deviation of the mass concentration $\underline{C} = \underline{\rho}_1/(\underline{\rho}_1 + \underline{\rho}_2)$ of the solute from its mean $\underline{C}_0$. Here $\underline{\rho}_1$ and $\underline{\rho}_2$ are the mass density fields of the two components. The thermal expansion coefficient $\alpha$ and the solutal expansion coefficient $\beta$ of the fluid come from a linear isobaric equation of state of the total mass density $\underline{\rho} = \underline{\rho}_1 + \underline{\rho}_2$,

$$\underline{\rho} = \underline{\rho}_0 [1 - \alpha(\underline{T} - \underline{T}_0) - \beta(\underline{C} - \underline{C}_0)], \qquad (2.5)$$

for small deviations from the mean values $\underline{T}_0$ and $\underline{C}_0$. For the room temperature solutions of ethanol in water that are commonly used in experiments $\alpha$ and $\beta$ are positive [3].

In the continuity equation (2.1), the fluid has been assumed to be incompressible, i.e., the mass density $\underline{\rho}$ is constant, and the mass current is proportional to the divergence-free velocity field $\mathbf{u}$. The reduced heat current $\mathbf{Q}$ in Oberbeck-Boussinesq approximation consists of the convective part $\mathbf{u}\, \delta T$ and the diffusive part $-\boldsymbol{\nabla}\, \delta T$. In the reduced concentration current $\mathbf{J}$, the diffusive part is $-L\boldsymbol{\nabla}\left(\delta C - \psi\, \delta T\right)$. Here, we suppress the convective transport $\mathbf{u}\, F_0$ of the mean quantity $F_0$, since it drops out in the balance equations. The Lewis number $L$ is the ratio of the concentration diffusivity $D$ to the thermal diffusivity $\kappa$, and the Prandtl number $\sigma$ is the ratio of the momentum diffusivity $\nu$ and $\kappa$:

$$L = \frac{D}{\kappa} \; ; \; \sigma = \frac{\nu}{\kappa}. \qquad (2.6)$$

For room temperatures ($10^o C - 40^o C$), the Prandtl number of ethanol-water mixtures lies between 5 and 20 [3], while for normal fluid Helium it is ten or more times smaller. The Lewis numbers are around 0.01 for these mixtures.

In experiments as well as in our simulations, the material parameters $L$ and $\sigma$ that characterize a particular mixture are fixed, while the Rayleigh number $R$ and the separation ratio $\psi$,

$$R = \frac{\alpha g d^3}{\kappa \nu}\Delta \underline{T} \; ; \; \psi = -\frac{\beta}{\alpha}\frac{k_T}{\underline{T}_0}, \qquad (2.7)$$

are considered to be control parameters that can be varied independently. $R$ measures the externally imposed thermal stress. The Soret coupling $\psi$ between temperature and concentration, into which enters the thermo diffusion ratio $k_T$ of the mixture, reflects the influence of temperature gradients on the concentration field. $\psi$ can be positive or negative [3], depending on $\underline{T}_0$ and $\underline{C}_0$. For room temperature ethanol-water mixtures, $\psi$-values between about $-0.5$ and $+0.2$ can be easily realized experimentally [3]. The Dufour cross-coupling term in the heat balance equation can be ignored for binary liquids, but it has to be kept for gas mixtures [84,65].

The buoyancy force $(\underline{\rho} - \underline{\rho}_0)\mathbf{g}$ due to density deviations from the mean is the driving mechanism for convective motion. It enters into the momentum balance (2.4) via the buoyancy term $\mathbf{B}$ which follows from (2.5) after scaling. This is the only place where a density variation is considered in Oberbeck-Boussinesq approximation.

Taking the divergence of the Navier Stokes equation (2.4), one gets, via the continuity equation, a Poisson equation

$$\nabla^2 p = -\boldsymbol{\nabla}\left[\boldsymbol{\nabla}\left(\mathbf{u} : \mathbf{u}\right)\right] + \sigma\, R\, \partial_z\, (\delta T + \delta C) \qquad (2.8)$$

for the pressure $p$. The Poisson equation substitutes the continuity equation and builds, together with (2.2–2.4), a complete set of equations for the fields $\mathbf{u}$, $\delta T$, $\delta C$, and $p$.

### B. Boundary conditions

The horizontal boundaries of the layer that we shall call plates for shortness are at $z = 0, 1$. The lateral boundaries are at $x = 0, \Gamma$. The plates are rigid for the fluid,

$$u = w = 0 \qquad \text{for } z = 0, 1, \tag{2.9}$$

and perfect heat conductors, so that the temperature of the fluid at $z = 0, 1$ is constant and laterally homogeneous:

$$\delta T = \frac{1}{2} \quad \text{at } z = 0, \quad \text{and}$$
$$\delta T = -\frac{1}{2} \quad \text{at } z = 1. \tag{2.10}$$

This is a good approximation for copper and sapphire plates that are used in most experiments, since their heat conductivity is very large compared to that of the fluid. The plates are impermeable to the fluid, so there is no concentration current through the plates:

$$\mathbf{J} \cdot \mathbf{e}_z = 0 \quad \text{or}$$
$$\partial_z \delta C = \psi \, \partial_z \delta T \qquad \text{at } z = 0, 1. \tag{2.11}$$

These NSI (no-slip impermeable) boundary conditions have to be contrasted with the idealized FSP (free-slip permeable) boundary conditions that allow expansion of the fields in trigonometric functions and that are often used in theoretical approaches. However, since the latter allow vertical concentration transport through the plates, they are likely to misrepresent the delicate concentration balance [46] that is involved, e.g., in explaining structure, dynamics, and stability of LTW states.

In lateral direction, all fields $F = u, w, T, C, p$ are periodic:

$$F(x, z; t) = F(x + \Gamma, z; t), \tag{2.12}$$

with a given lateral periodicity length $\Gamma$. Since the pressure $p$ is determined via the Poisson equation (2.8) by $\mathbf{u}, \delta T, \delta C$, we do not need boundary conditions for it.

Most of the nonlinear states determined in this paper were obtained in a periodicity interval of $\Gamma = \lambda = 2$. Thus the wave number is fixed at $k = \pi$, which is close to those observed in experiments.

### C. Numerical method

To integrate the partial differential equations, we used a modification of the SOLA code that is based on the MAC method [85–88]. This is a finite-difference method of second order in space on staggered grids for the different fields, with a uniform spatial resolution $\Delta x = \Delta z = 0.05$ and with an explicit first-order Euler step in time. The pressure field was iteratively calculated from the Poisson equation (2.8) using the artificial viscosity method [88]. Most calculations were done on a CRAY. For more details see [49].

### D. Conductive state

The equations of sec. II.A together with the boundary conditions of sec. II.B have a stationary, laterally homogeneous solution, for all parameters

$$\mathbf{u}_{cond} = \mathbf{0} \quad ; \quad \delta T_{cond} = \frac{1}{2} - z \; ;$$
$$\delta C_{cond} = \psi \left(\frac{1}{2} - z\right), \tag{2.13}$$

which is globally stable for small thermal stress. It describes vertical heat diffusion through the layer without convective motion. The heat current is $\mathbf{Q}_{cond} = \mathbf{e}_z$. The constant temperature gradient induces a vertical concentration gradient (given by $-\psi$) such that the concentration current vanishes, $\mathbf{J}_{cond} = \mathbf{0}$, everywhere – not only at the plates. The concentration stratification causes in the buoyancy

$$\mathbf{B}_{cond} = \sigma R \, (1 + \psi) \left(\frac{1}{2} - z\right) \mathbf{e}_z \tag{2.14}$$

a modification by the factor $(1 + \psi)$ relative to the thermal part. So, depending on the sign of the Soret coupling, the Soret effect enhances or depresses the thermal part of the buoyancy. The pressure is $p_{cond} = -\frac{1}{2} \sigma R \, (1 + \psi) \left(\frac{1}{2} - z\right)^2$.

To describe convection in the following, we sometimes use the deviations of the fields from the conductive state,

$$\theta = \delta T - \delta T_{cond} \; ; \; c = \delta C - \delta C_{cond}, \tag{2.15}$$

and similarly for the buoyancy force

$$\mathbf{b} = \mathbf{B} - \mathbf{B}_{cond} = b \, \mathbf{e}_z \; ; \; b = \theta + c. \tag{2.16}$$

### E. Soret coupling

The Soret effect causes a linear coupling between the degrees of freedom of the concentration field on the one hand and those of the temperature and also of the velocity field on the other hand. It is this coupling to the concentration field that causes the additional richness of convective dynamics and structures in binary mixtures in comparison to pure-fluid convection. On the one hand, the concentration field is mixed and advected passively by the nonlinear $\mathbf{u} \, \delta C$ coupling. But, on the other hand, it enters via the Soret effect linearly into the buoyancy force that drives convection. Thus, the structural dynamics of the concentration field actively feeds back into the driving mechanism for convection. Without Soret coupling, $\psi = 0$, the temperature gradient of the conductive state does not induce a vertical concentration gradient – any concentration deviation from the mean diffuses away. And since convection does not create new concentration fluctuations, the buoyancy force is not modified. So, in the limiting case of zero coupling $\psi = 0$, the equations effectively describe a pure fluid. For $\psi \neq 0$, however, the externally-imposed temperature stress sustains — against the mixing action of convection — via the Soret effect concentration variations which in turn alter the driving force for convection.

### F. Bifurcations out of the conductive state

The stability properties of the conductive state against infinitesimal convective perturbations have been discussed in detail in the literature [66–69]. For ethanol-water parameters, $L = 0.01$ and $\sigma = 10$, the results are summarized in Fig. 1 as a function of $\psi$ for the experimentally accessible $\psi$-range.

Here and in the rest of this paper we use the scaled Rayleigh number

$$r = \frac{R}{R_c^0} \qquad (2.17)$$

that is reduced by the critical Rayleigh number $R_c^0$ for onset of convection in a pure fluid with the critical wave number $k_c^0$. The analytical values are $R_c^0 = 1707.762$ and $k_c^0 = 3.11632$. They are used in Fig. 1 for the reduced critical values of $r$ and of $\widehat{k} = k/k_c^0$. In the finite-differences approximation of the field equations that is used in our MAC algorithm, however, the threshold for onset of pure fluid convection [89] lies at

$$R_c^0 = 1686(\pm 0.2\%)$$
$$\text{for} \qquad k = \pi \; ; \quad \Delta x = \Delta z = \frac{1}{20} \, . \qquad (2.18)$$

So, when presenting numerical results obtained with the above discretization, we scale Rayleigh numbers by (2.18).

The lines $r_{stat}$ and $r_{osc}$ in Fig. 1a are bifurcation thresholds where convective solutions branch out of the conductive state, and Fig. 1b shows the critical reduced wave numbers there. At the full and dash-dotted line in Fig. 1a, a SOC solution branches off the ground state. For $\psi$ above (below) the tricritical value $\psi_{SOC}^t = -O(10^{-7})$ [69], the SOC bifurcation is forwards (backwards). However, below $\psi_{SOC}^\infty = -L/(1 + L)$ [36,65] where $r_{stat}$ diverges, the lower SOC solution branch is disconnected for positive $r$ from the conductive solution [71].

The dashed line in Fig. 1a denotes a Hopf bifurcation threshold where symmetry-degenerated left and right TW solutions branch out of the conductive state. The Hopf oscillation frequency at onset, $\omega_H$, is shown in Fig. 1c. It varies as $\omega_H^2 \approx -449 \, \psi/(1 + \psi + 1/\sigma)$ [27,18]. For $\psi$ above (below) the tricritical value $\psi_{TW}^t = -O(10^{-4})$ [69,70] the TW bifurcation is forwards (backwards). The critical bifurcation thresholds $r_{stat}$ and $r_{osc}$ intersect in Fig. 1a at the codimension-two value $\psi_{CTP} = -3.5 \cdot 10^{-5}$ with slightly different critical wave numbers and a small Hopf frequency. For a more detailed discussion of the codimension-two point see, e.g., [66–69,72,73].

The linear quantities of Fig. 1 do not change much as long as $\sigma$ remains significantly bigger and $L$ significantly smaller than 1. For a more detailed discussion of the $L$, $\sigma$ dependence, cf. sec. VI.

### G. Order parameters

To characterize the convective solutions we use different order parameters. (i) The maximal vertical velocity $w_{max}$ directly measures the convective amplitude. (ii) The Nusselt number

$$N = \frac{1}{\Gamma} \int_0^\Gamma dx \, Q_z \qquad (2.19)$$

is the total vertical heat current through the fluid layer, $\int_0^\Gamma dx \, Q_z$, reduced by the conductive part, $\int_0^\Gamma dx \, Q_{cond} = \Gamma$. In our scaling, $Q_{cond} = 1$. We evaluate $N$ at $z = 0$ and in addition at $z = 1$ to check that our code does not violate energy conservation. The reduced vertical heat current carried by convection alone, $N - 1$, measures squared convective amplitudes of velocity or temperature fields.

(iii) Since $w_{max}$ and $N - 1$ do not characterize the concentration field in the nonlinear convective states appropriately, we introduce a quantity that characterizes the magnitude of concentration variations. To that end, we use the mixing parameter

$$M = \sqrt{\langle \delta C^2 \rangle \Big/ \overline{\delta C_{cond}^2}} \, . \qquad (2.20)$$

$M$ is the variance of the concentration field reduced by its value in the conductive state. Here brackets and overbar imply lateral and vertical averages, i.e., a spatial average over the whole fluid volume. In a perfectly mixed mixture where all concentration deviations $\delta C$ from the mean vanish, $M$ would be zero. On the other hand, in the conductive state with the Soret-induced concentration gradient, $M$ is defined to be 1. So $1 - M$ is an order parameter for the convective state that is zero in the ground state and approaches $+1$ for convection with strong mixing properties.

(iv) TW states are characterized by their oscillation frequency $\omega$.

### III. NEGATIVE SORET COUPLING

In this section, we discuss the structure and bifurcation behavior of extended convective states in mixtures with buoyancy-reducing negative Soret effect. We start with a typical bifurcation diagram for an ethanol-water mixture. We discuss the structure of the fields of the TW and SOC states in this diagram, including symmetry properties, lateral Fourier decomposition, and of time averaged currents of mass, heat, and concentration. The variation of bifurcation behavior and field structure with separation ratio, for the case of a buoyancy-enhancing positive Soret effect, and the influence of changing the Lewis and Prandtl number on the bifurcation structure are presented in Sections IV, V, and VI, respectively.

### A. Bifurcation behavior

Here we discuss bifurcation diagrams of extended states in a mixture with parameters $L = 0.01$, $\sigma = 10$, $\psi = -0.25$. Fig. 2 shows the $r$-dependence of frequency $\omega$, mixing parameter $M$, and Nusselt number $N - 1$ of the convective states with wavelength $\lambda = 2$. These diagrams display in a representative way the subcritical bifurcation topology that is typical for sufficiently negative Soret coupling $\psi$. The dotted line gives for comparison the Nusselt number of the SOC states in a pure fluid. The conductive state loses its stability at $r_{osc}$. Just above $r_{osc}$, the system does not saturate in a state with small convective amplitude as in a pure fluid; rather, a nonlinear TW state of large amplitude is stable. This TW state has a frequency of only about $\frac{1}{10}\omega_H$. Reducing the drive $r$ quasistatically, the TW frequency continuously increases up to about $\frac{1}{3}\omega_H$ at the saddle $r_{TW}^s$. Below the saddle, there is no longer a TW branch. So, with further reduction of the drive below $r_{TW}^s$, the system undergoes a transition to the conductive state. The arrows in Fig. 2c indicate the hysteretic character of the bifurcation. The reason for the associated discontinuous onset of convection is a nonlinear feedback: For $\psi < 0$, the Soret-induced conductive concentration distribution weakens the buoyancy. Convection, on the other hand, redistributes the alcohol more evenly, thereby reducing the adverse Soret effect and thus increasing the buoyancy much more strongly than the flattening of the vertical temperature profile in the bulk decreases the buoyancy. The increased buoyancy in turn strengthens convection, which again amplifies the buoyancy.

With increasing $r$, the TW frequency decreases until there is a continuous transition, at $r^*$, to a TW of frequency zero, i.e., a SOC state. Approaching $r^*$, the TW frequency decreases as $(r^* - r)^{\frac{1}{2}}$. From a fit $\omega(r) = a(r^* - r)^b$ with parameters $a$, $r^*$, and $b$, we got from our data $a = 0.4\,\pi \pm 15\%$, $r^* - 1 = 0.65 \pm 0.5\%$, $b = 0.5 \pm 10\%$. Ohlsen et al. [11] obtained a slightly different exponent $b = 0.58$ from their experiments. Bensimon et al. [51] got $b = 0.5$ from their concentration-boundary-layer model. For larger $r$, we obtained stable SOC states. The transition at $r^*$ between the upper TW branch and upper SOC branch (squares in Fig. 2) is smooth and non-hysteretic.

Viewed from the SOC branch above $r^*$, we have, with decreasing $r$, a symmetry-breaking forward bifurcation to left- or right-traveling waves, the symmetry of the latter being reduced relative to that of the former (cf. sec. III.C). A similar bifurcation with $\omega$ as order parameter has been discussed in another connection, cf. eg. [90,91].

It is possible to stabilize the SOC branch below $r^*$ (triangles in Fig. 2) down to the SOC saddle $r_{SOC}^s$ by preventing this symmetry breaking. In most cases, we enforced the stability of the otherwise unstable SOC's by hindering the traveling of the rolls with the phase pinning boundary condition $u(x, z; t) = 0$ at $x = 0, \Gamma$ in addition to the periodic boundary conditions in $x$ direction. This means that we do not allow lateral convective transport through the line $x = 0$. So, in our short system of $\Gamma = 2$, the roll pair cannot travel. In somewhat longer systems, this boundary condition is not sufficient to prevent TW's in the bulk, as has been seen in simulations of a system of length $\Gamma = 10$ [49] and similarly in experiments [4,28–30].

The Nusselt numbers of TW and SOC states typically lie just below the values for a pure fluid. Only near the saddles are there stronger deviations. The stabilized SOC's transport more heat vertically than the corresponding TW states for the same parameters. We should mention here that the conductive state is also with phase pinning boundary conditions stable only up to $r_{osc}$. A transient standing wave of small amplitude grows above $r_{osc}$. However, a final nonlinear standing-wave state is not stabilized by the phase pinning boundary conditions. Instead, the system ends in the SOC state.

If we start from a stabilized SOC state (triangles in Fig. 2) below $r^*$ and then release the phase pinning, the SOC state becomes unstable. The convection rolls start to travel, and with growing phase velocity the Nusselt number decreases. Finally, for $r_{TW}^s < r < r^*$, the system ends in the stable TW state. For $r < r_{TW}^s$, on the other hand, there is no stable convective state. Below $r_{TW}^s$, the convection amplitude decreases, while the phase velocity increases, until finally the system approaches the conductive state as a linear TW of exponentially-decreasing amplitude.

That our SOC and TW states are closer to SOC states in a pure fluid than to the conductive state of the mixture is reflected not only in the Nusselt number, but also in the mixing parameter $M$ (2.20) (cf. Fig. 2b) of the concentration field. $M$ is normalized to 1 in the conductive state with the Soret-induced vertical concentration variation. Convective mixing dramatically reduces this concentration variation to $M \approx 0.1$, with a slight increase at the saddle. There is a stronger variation of the concentration field in the TW's. There, $M$ varies roughly linearly with frequency $\omega$, for not too small $\omega$, and reaches its maximum ($\approx 0.35$) at the saddle $r_{TW}^s$. Thus, the convective homogeneization of the concentration, measured in terms of $M$, is more efficient by a factor of 3.5 in SOC's than in a TW at the saddle. A similarly-defined mixing parameter of the temperature field is reduced to only 0.9 for the strongest convective state in Fig. 2.

The bifurcation diagrams in Fig. 2 were obtained in a system with periodicity length $\Gamma = \lambda = 2$. Thus, some of the instability mechanisms, e.g., the Eckhaus instability [31–33] and more complicated dynamical behaviour [10,34,48,35,49,1,46, and references cited therein] occurring in large systems are suppressed here.

## B. Structure of the fields

Here and in the next sections we discuss how the fields of the nonlinear convective states change along the upper bifurcation branch. To that end we display in Fig. 3 three TW states and one SOC state contained in the bifurcation diagrams Fig. 2. Their symmetry properties will be discussed in sec. III.C, while vertical profiles and a lateral Fourier analysis are presented in Fig. 4 and sec. III.D.

Let us first discuss the fast TW close to the saddle in Fig. 3a. This TW travels with phase velocity $v_p = \omega/k$ to the right. Since all its constituent fields have the form

$$\delta F(x, z; t) = \delta F(x - v_p t, z; 0), \qquad (3.1)$$

it is a stationary structure with the given spatial periodicity

$$\delta F(x + \lambda, z; t) = \delta F(x, z; t) \qquad (3.2)$$

with $\lambda = 2$. The region $-1.275 < x < 1.225$ of Fig. 3a covers two and a half convection rolls, and the maximum of the vertical velocity is at $x = 0$. The first row shows the colour-coded temperature field $\delta T$ together with arrows representing the velocity field $\mathbf{u}$. Close to the maximal upflow at $x = 0$, the isotherms are bent upwards, because warm fluid is transported upwards from the warmer lower plate there. Correspondingly, the isotherms are bent downwards close to the maximal downflow at $x = \pm 1$. Both velocity and temperature fields look like those in a pure fluid. But here in the mixture, the temperature field is slightly phase shifted relative to the vertical velocity field. The phase shift can be seen most clearly by comparing the lateral wave profiles at $z = 0.5$ in the fourth row of Fig. 3. They show that the temperature wave $\delta T$ (thin line with triangles) lags behind the wave of the vertical velocity $w$ (thin line). The phase shift between $w$ and $T$ grows monotonically with frequency [57], roughly like $\arctan(\omega/(k^2+\pi^2))$ [55], and is largest for linear TW's [68] at the Hopf bifurcation threshold $r_{osc}$. The temperature wave is not planar: in fact, the lines of constant phase of the first lateral Fourier modes of all TW fields are curved [57] to a varying degree instead of being straight, vertical lines.

The third row contains the colour-coded concentration field $\delta C$ together with the streamlines of the velocity field in a frame comoving with the TW. While the velocity and temperature fields are laterally nearly harmonic, the lateral concentration profile has a trapezoidal, strongly nonharmonic shape — see the thick line with squares for $\delta C(x, z = 0.5)$ in the fourth row of Fig. 3a. For a TW propagating to the right, $\delta C$ has a positive plateau with a concentration surplus (blue-green area) in the region of the right turning convection rolls and correspondingly a negative plateau with a concentration deficit (red-orange area) in the region of the left turning rolls. Between the two concentration plateaus there is an almost linear concentration variation in horizontal direction. Also, from a plateau to a plate there is a steep, almost linear variation of $\delta C$ (cf. Fig. 4c).

To explain this structure of the TW's concentration field self-consistently, we view the fluid layer from the frame of reference $\widetilde{\Sigma}$ that comoves with the phase velocity $v_p \mathbf{e}_x$ of the TW. In this frame, all fields $\delta \widetilde{F}$ of the stationary TW

$$\widetilde{\mathbf{u}}(\widetilde{x}, z) = \mathbf{u}(x - v_p t, z; t) - v_p \mathbf{e}_x \qquad (3.3)$$

$$\delta \widetilde{T}(\widetilde{x}, z) = \delta T(x - v_p t, z; t) \qquad (3.4)$$

$$\delta \widetilde{C}(\widetilde{x}, z) = \delta C(x - v_p t, z; t) \qquad (3.5)$$

are time-independent, with

$$\widetilde{x} = x - v_p t. \qquad (3.6)$$

From the conservation law

$$\partial_t \delta \widetilde{C} + \widetilde{\boldsymbol{\nabla}} \cdot \widetilde{\mathbf{J}} = 0 \qquad (3.7)$$

and $\partial_t \delta \widetilde{C} = 0$, it follows that the concentration current

$$\widetilde{\mathbf{J}} = \widetilde{\mathbf{u}} \, \delta \widetilde{C} - L \, \widetilde{\boldsymbol{\nabla}} \left( \delta \widetilde{C} - \psi \, \delta \widetilde{T} \right) \qquad (3.8)$$

is divergence-free. Then the continuity equation (2.1) implies

$$\left( \widetilde{\mathbf{u}} \cdot \widetilde{\boldsymbol{\nabla}} \right) \delta \widetilde{C} = L \, \widetilde{\nabla}^2 \left( \delta \widetilde{C} - \psi \, \delta \widetilde{T} \right). \qquad (3.9)$$

Since the Lewis number $L$ is small, we can set the diffusive term on the r.h.s. of (3.9) to zero for the present discussion. Thus, there is no gradient of $\delta \widetilde{C}$ parallel to $\widetilde{\mathbf{u}}$; i.e., the isolines of $\delta \widetilde{C}$ lie practically parallel to $\widetilde{\mathbf{u}}$. Furthermore, if we define the velocity streamfunction $\widetilde{\Phi}$ in $\widetilde{\Sigma}$ via

$$-\widetilde{\boldsymbol{\nabla}} \times \widetilde{\Phi} \mathbf{e}_y = (\partial_z \widetilde{\Phi}, 0, -\partial_x \widetilde{\Phi}) = (\widetilde{u}, 0, \widetilde{w}), \qquad (3.10)$$

then the isolines of $\widetilde{\Phi}$, i.e., the streamlines of $\widetilde{\mathbf{u}}$ along which the fluid moves in $\widetilde{\Sigma}$, are almost identical with the isolines of the concentration. That this holds quite well can be seen in Fig. 3 by comparing the colour-coded concentration distributions with the dashed superimposed streamlines of $\widetilde{\mathbf{u}}$. So, in order to understand the structure of the concentration field, we first investigate the form of the streamlines. The latter can be most easily understood within a single-mode representation [55],

$$\widetilde{\Phi} = -\frac{1}{k}\,\widetilde{w}(z)\,\sin(k\widetilde{x}) - v_p\,z\,. \qquad (3.11)$$

This streamfunction captures the characteristic properties of the velocity field being basically harmonic in the lateral direction with a vertical profile $\widetilde{w}(z)$. The second contribution, $-v_p z$, to (3.11), that arises in $\widetilde{\Sigma}$ from the phase velocity of the TW, has two effects: (i) it shifts the closed isolines of the ordinary first contribution to (3.11) — corresponding to stationary roll convection — alternatingly towards the top and bottom plates, and (ii) it generates open streamlines. These meander between and around the roll-like regions of closed streamlines (Fig. 3a), whereas in an SOC ($v_p = 0$), the only open streamlines are the vertical separation lines between the oppositely turning rolls joining top and bottom plate. Consequently, in a right-propagating TW, (i), the blue-green, concentration rich (red-orange, concentration poor) regions of closed streamlines for the right (left) turning fluid domains in $\widetilde{\Sigma}$ are displaced towards the upper cold (lower warm) plate, where the Soret effect has caused a concentration surplus (deficiency) for $\psi < 0$. In addition, (ii), open streamlines separate the blue-green (red-orange) regions from the bottom low-concentration (top high-concentration) boundary layer. Therefore, the blue-green (red-orange) boundary layers feed high concentration (low concentration) only into the respective rolls. See the downwards blue (upwards red) jet-like structures of the concentration field pointing into the two regions of closed streamlines in Fig. 3a. Within the regions of closed streamlines, the concentration field behaves like a passive scalar, mixing up diffusively [78] as the fluid is advected around in $\widetilde{\Sigma}$ along the closed streamlines. Since a volume element in the area of closed streamlines moves on a closed orbit in $\widetilde{\Sigma}$, the fluid in such an area is caught within it if we ignore diffusion. Thus, in the laboratory frame, this entire fluid volume is transported with the phase velocity $v_p\,\mathbf{e}_x$ of the TW to the right. So the concentration wave moves about $\frac{1}{4}$ wavelength ahead of the $w$-wave (see the profiles of the fields in the 4th and 5th row of Fig. 3).

The above explanation of the concentration field structure is based on the smallness of $L$ and on the fact that $\delta C$ shows some of the characteristic behaviour of a passively convected field. If one tried to explain the structure of the temperature field along similar lines, starting from the divergence freeness, $\widetilde{\boldsymbol{\nabla}} \cdot \widetilde{\mathbf{Q}} = 0$, of the heat current

$$\widetilde{\mathbf{Q}} = \widetilde{\mathbf{u}}\,\delta\widetilde{T} - \widetilde{\boldsymbol{\nabla}}\,\delta\widetilde{T} \qquad (3.12)$$

in $\widetilde{\Sigma}$, one would see that in this case (for $|\mathbf{u}|$ of the order of 10) the diffusive term in

$$\left(\widetilde{\mathbf{u}} \cdot \widetilde{\boldsymbol{\nabla}}\right) \delta\widetilde{T} - \widetilde{\nabla}^2\,\delta\widetilde{T} = 0 \qquad (3.13)$$

*cannot* be assumed to be small compared to the convective term as in the concentration case.

Let us now discuss the TW variation with decreasing frequency, i.e., increasing $r$. Since the phase shift between velocity and temperature wave varies like $\arctan(\omega/(k^2 + \pi^2))$ [55,57], it vanishes almost linearly with $\omega$ at $r^*$. From the streamfunction $\widetilde{\Phi}$ (3.11), one furthermore infers that, with decreasing phase velocity $v_p$ of the TW, (i) the alternating up- and downwards shift of the closed streamline regions and (ii) the transverse extension of open streamlines of $\widetilde{\mathbf{u}}$ between the rolls as well as between a roll and a plate decreases. Both effects lead to a reduction of the concentration contrast between the differently displaced and rotating regions of closed streamlines: decreasing $v_p$ makes the streamline structures more and more mirror symmetric around $x = 0$. Secondly, the feeding of concentration surplus (deficiency) from the top (bottom) boundary layer into the different roll regions becomes more symmetric.

Let us now try to understand the structure of the slow TW's and the SOC states. The picture of mixing of the TW concentration field within a region of closed streamlines of $\widetilde{\mathbf{u}}$ was based on the assumption that the convective term in the concentration balance dominates. For small phase velocities, and especially for SOC states, this picture has to be extended, since these closed streamlines cover an entire convection roll (see third row in Fig. 3d). Since the velocity field vanishes at the plates, concentration diffusion has "long enough time" to work before a fluid volume is transported away from the plates by the velocity field, and a Soret-induced concentration gradient develops at the plates. The turnover time of a fluid element that starts very close to a plate on a closed streamline is indeed much bigger than the vertical concentration diffusion time $\frac{1}{L}$. Since the spatial area where the turnover time is at least comparable to $\frac{1}{L}$ is small, the concentration boundary layer at the plates is small. Since streamlines through the top layer also pass through the bottom boundary layer, a fluid element on such a streamline experiences a strong change of its concentration content. Let us now follow a volume element starting close to the upper plate, say, at $x = 0.5$ with positive $\delta C$ and negative $\delta T$. First, it moves slowly with decreasing velocity to the right roll boundary while retaining its temperature and concentration. After that, it moves slowly away from the upper plate, since there $w$ is just proportional to the square of the distance from the plate. Then it is transported very quickly downwards through the middle of the layer, where $w_{\max}$ is of the order of 10. Since heat diffusion works quickly, the element warms up rapidly, and the isotherms there are bent only moderately downwards, as in a pure fluid. However, with concentration diffusion working very slowly on the timescale of $\frac{1}{L} = 100$ thermal diffusion times, the fluid element retains positive concentration down to the vicinity of the lower plate. This causes the little peak in the lateral profile of $\delta C$ in Fig. 3d at the position $x = \pm 1, z = 0.5$ of maximal downflow. In the vicinity of the lower plate, the volume element is transported very slowly further to the lower plate and then to the left. In this part of the cycle concentration diffusion again has enough time to work. Surplus concentration is now transported diffusively – mostly upwards – out of the volume element, and the concentration content of the fluid volume becomes negative. All these effects together generate the typical plume structure of the SOC concentration field with the shafts of the plumes centered at the positions of maximal up- and downflow. Such structures have also been seen in simulations of thermohaline convection [74,75]. The concentration plumes have rather narrow vertical extension and modify the concentration field in the interior of the rolls only very slightly. Similar plumes occur in the temperature field in very high Rayleigh number convection in a pure fluid, as a consequence of the much faster convective transport of heat compared to heat diffusion, where Veronis [92] has also seen them.

While the plumes, or concentration jets, in the SOC are mirror symmetric around $x = 0$, this is not the case in TW's. In particular, in the fast TW of Fig. 3a, the concentration jets alternatingly feed only the left or right turning rolls. There, the plumes are strongly bent into the respective rolls, and the resulting fine structure of concentration enhancement (depletion) causes the peak (dip) at $x \simeq 1$ ($x \simeq 0$). Since the turnover time also diverges [55,79] on the separatrix of the closed streamlines in a TW, this peak (dip) fine structure on top of the smooth background of the interior of the rolls thus has the same origin in TW's as in SOC's.

### C. Symmetry properties

The concentration field most clearly shows that SOC and TW states discussed here are strongly nonlinear structures. Nevertheless, they have, in addition to their stationarity (3.1) and spatial periodicity (3.2), an additional mirror-glide symmetry

$$\delta F(x, z; t) = \pm \delta F\left(x + \frac{\lambda}{2}, 1 - z; t\right) \qquad (3.14)$$

with + for $u$, $p$ and − for $w$, $\delta T$, $\delta C$ and $B$. This is a combination of a translation by half a wavelength in the $x$ direction and a reflection through the $z = \frac{1}{2}$ plane. Besides (3.1) and (3.2), this is the most elementary, i.e., longest persistent symmetry of SOC's and TW's. As far as we know, the mirror-glide symmetry (3.14) has first been discussed in ref. [54]. Veronis found that Fourier modes breaking this symmetry decayed to zero for thermohaline convection with fixed spatial phase (i.e. no TW was possible) [74] and for SOC states in one-component fluids [92]. But the symmetry was not explicitly stated. The mirror-glide symmetry of SOC rolls was recently verified numerically in convection of one-component fluids [89] as well.

The symmetry (3.14) can easily be seen in the light intensity distribution

$$I(x, z) = A \nabla^2 \left[\delta C(x, z) + b \, \delta T(x, z)\right]. \qquad (3.15)$$

that one would observe in side-view shadowgraphs obtained by shining light horizontally through the fluid layer along the roll axes [93,94,3,14,15]. $A$ is a constant, and $b = -0.919$ with our parameters for a real mixture [14,3]. Fig. 3 shows this light intensity distribution in grey scale. Note that its characteristic structures are dominated by the large Laplacians of the concentration field. This incidentally shows that the shadowgraph method also gives information about the concentration field [14,15].

Applying the operation (3.14) twice yields the pure spatial translation symmetry (3.2) by one wavelength. On the other hand, if one combines (3.14) with the stationarity (3.1), one gets a space-time symmetry

$$\delta F(x, z; t) = \pm \delta F\left(x, 1 - z; t + \frac{\tau}{2}\right), \qquad (3.16)$$

with $\tau = 2\pi/\omega$ for the TW states. Here the translation in (3.14) by half a wavelength is replaced by a time translation by half a period. This space-time symmetry was mentioned in another connection by Weiss [95]. Note that (3.16) does not imply a horizontal mirror symmetry at the midplane for SOC states – which they do not have anyhow – since $\tau$ diverges for $\omega = 0$. But in a system $\check{\Sigma}$, that moves with constant velocity $\check{V}\mathbf{e}_x$ relative to the laboratory system, the spatial symmetry (3.1) becomes a space time symmetry

$$\delta \check{F}(\check{x}, z; t) = \pm \delta \check{F}\left(\check{x}, 1 - z; t + \frac{\check{\tau}}{2}\right) \qquad (3.17)$$

with $\check{\tau} = \lambda/\check{V}$. This also holds for TW's with $\check{V}$ being the velocity of $\check{\Sigma}$ relative to the rest system $\tilde{\Sigma}$ of the TW.

The only additional symmetry of SOC's compared to TW's is the mirror symmetry

$$\delta F(x, z) = \pm \delta F(-x, z) \qquad (3.18)$$

with − for $u$ and + for $w$, $\delta T$, $\delta C$, $p$, and $B$, where $x = 0$ is the position of a roll boundary. All other symmetries of the SOC states follow from (3.1), (3.14), (3.18). For example, the point mirror symmetry around the roll center $x = \frac{\lambda}{2}$, $z = \frac{1}{2}$,

$$\delta F(x, z) = \pm \delta F\left(\frac{\lambda}{2} - x, 1 - z\right), \qquad (3.19)$$

with + for $p$ and − for $u$, $w$, $\delta T$, $\delta C$, and $B$, is a combination of (3.14) and (3.18). Finally, we mention that linear SOC and TW *transients* have no further symmetry besides those discussed here.

### D. Vertical profiles and lateral Fourier decomposition

In this section, we continue to discuss how the structure of convection changes with frequency and Rayleigh number. To that end, vertical profiles of the concentration field at different lateral positions are shown in Fig. 4c, and the profiles of lateral Fourier modes of the temperature and concentration field are shown in Fig. 4a and Fig. 4b, respectively, for the three TW states and the SOC state in Fig. 3. The Fourier decomposition of the fields for these states is

$$\delta F(x,z;t) = \delta \widehat{F}_0(z) + Re \sum_{n=1}^{\infty} \delta \widehat{F}_n(z;t)\, e^{i n k x} \quad (3.20)$$

with $k = \frac{2\pi}{\lambda}$. Because of (3.1), the moduli of the Fourier coefficients,

$$\delta \widehat{F}_n(z;t) = |\delta \widehat{F}_n(z)|\, e^{-i\, \varphi_{F_n}(z;t)}, \quad (3.21)$$

are time independent, while in a TW the phases $\varphi_{F_n}$ vary linearly in time. Fig. 4a shows the vertical variation of the Fourier modes $|\widehat{\theta}_n(z)| = |\delta \widehat{T}_n(z)|$ of the temperature field for $n = 1, 2, 3$ and of $\widehat{\theta}_0(z) = \delta \widehat{T}_0(z) - \left(\frac{1}{2} - z\right)$, i.e., the laterally averaged deviation from the conductive profile.

Due to nonlinear effects, the profile of $|\widehat{\theta}_1|$ flattens as convection intensifies with growing $r$. The nonlinearly stimulated higher harmonics $|\widehat{\theta}_2|, |\widehat{\theta}_3|$ grow with $r$ but are weak over the $r$-region considered here. So, as in a pure fluid [83,89], the temperature field of SOC's and TW's can be accurately approximated by two lateral Fourier modes

$$\delta T(x,z;t) \simeq \delta \widehat{T}_0(z) + |\delta \widehat{T}_1(z)|\, \cos(k\,x - \varphi_{T_1}(z;t))\,.$$
$$(3.22)$$

For a TW, the phase profile

$$\varphi_{T_1}(z;t) = \varphi_{T_1}(z;t=0) + k\, v_p\, t \quad (3.23)$$

is $z$-dependent [54], while, in a SOC, $\varphi_{T_1}$ is constant in $z$ and $t$. The vertical derivative of the mean deviation $\widehat{\theta}_0$ from the conductive temperature field at $z = 0, 1$ gives the convective heat current. All in all, the temperature field is similar to that in a pure fluid [89]. This also holds for velocity modes which are not displayed in Fig. 4.

The behavior of the concentration field is drastically different. It shows four characteristic features: (i) convective reduction of the Soret-induced difference between top and bottom plates, (ii) homogeneous distribution in the roll-like regions of closed streamlines of Fig. 3, (iii) linear boundary-layer variation, and (iv) plumes. Already close to the TW saddle, the concentration difference between the plates, $C(z=1) - C(z=0)$, is convectively reduced to about $\frac{1}{3}$ of the value for the conductive profile (dotted line in Fig. 4c). This ratio is further reduced to $\frac{1}{6}$ for the SOC state in the right column. The reason is that impermeability [27] imposes a condition only on the vertical derivative of the concentration field at the plates, while the temperature is fixed via a Dirichlet condition at the plates. A similar reduction of the concentration difference between the plates has also been found in the Galerkin model of Linz et al. [71,55] for impermeable boundary conditions. The concentration current at the plates,

$$\mathbf{J}(x,z;t) = -L\, \partial_x\, \delta C(x,z;t)\, \mathbf{e}_x \qquad \text{for } z = 0, 1\,, \quad (3.24)$$

is horizontal and small, but in general nonzero.

The homogeneous distribution in the roll-like regions in Fig. 3 shows up in Fig. 4c as plateaus of the vertical profiles of $\delta C$. Note that the profiles A, B are related to profiles A', B' via the symmetry (3.14). Curve A (A') refers to a section through a right (left) turning roll at $x_A = 0.35$ ($x_{A'} = -0.65$) close to the roll center at $\frac{1}{2}$ ($-\frac{1}{2}$). Profile B at $x_B = 0.85$ (B' at $x_{B'} = -0.15$) is slightly further away from the center, so that the vertical extension of the plateau region is slightly smaller there. Since $x_B$ ($x_{B'}$) is close to a downflow (upflow) jet from the concentration rich top (poor bottom) plate, curve B (B') lies in the upper (lower) part of the layer above (below) curve A (A'). The high concentration blue (low concentration red) plume that is best visible in the TW of Fig. 3a leads to a concentration enhancement (depletion) in the low-$z$ (high-$z$) part of the vertical profile A (A'). This consequence of the previously-described plumes can also be seen in the profiles of the slow TW's and SOC of Fig. 4c.

Since the convective mixing of concentration is very effective, the laterally-averaged concentration $\langle \delta C \rangle(z)$ is nearly zero over a wide range in the bulk of the layer. For our stationary TW's and SOC's, $\langle \bullet \rangle$ denotes a lateral average over a wavelength

$$\langle f \rangle(z) = \frac{1}{\lambda} \int_0^\lambda f(x,z;t)\, dx\,, \quad (3.25)$$

which for TW's is equivalent to a *time* average over a period

$$\langle f \rangle(z) = \frac{1}{\tau} \int_0^\tau f(x,z;t)\, dt\,. \quad (3.26)$$

The impermeable boundaries cause a vertical concentration gradient

$$\partial_z \langle \delta C \rangle(z) = -\psi\, N \qquad \text{for } z = 0, 1 \quad (3.27)$$

at the plates. Although it is larger than the gradient, $\partial_z C_{cond} = -\psi$, in the conductive state (dotted line in Fig. 4c), the *overall* concentration difference between the plates is strongly reduced. In this way, the density difference between the plates and with it the buoyancy force in the fluid has increased, as compared with the conductive state. For Rayleigh numbers larger than those presented here, mixing effects can also be seen in the temperature field. Together with those of the concentration, they lead to a mixing in the mass density, so that there is no vertical gradient of the mass density in part of the area of closed streamlines (cf. Fig. 3b of [54]).

As a consequence of the reduction in the concentration difference between adjacent rolls with decreasing TW frequency, the strength of the concentration Fourier modes decreases, in particular in the middle of the cell. Since the area of the open streamlines of $\widetilde{\mathbf{u}}$ also shrinks, the plateaus of $|\delta \widehat{C}_1|$ broaden (thick lines in Fig. 4b). In the SOC state, $\delta C$ nearly vanishes in the middle of the cell, as can already be inferred from the smallness of the mixing parameter $M$ (Fig. 2b). Only the boundary layers between the rolls and the plates remain. However, the lateral variation of the concentration right at the plates and in its immediate vicinity — see the little orange (blue) dip at the origin of the upwards jet at $x = 0$ (downwards jet at $x = \pm 1$) in Fig. 3d — implies that $\delta \widehat{C}_1$ is finite at $z = 0, 1$ (cf. Fig. 4b). In fact the SOC concentration right at the plates obeys a Laplace equation

$$\left(\partial_x^2 + \partial_z^2\right) \delta C(x, z) = 0 \qquad \text{for } z = 0, 1. \qquad (3.28)$$

The boundary-layer behavior of the concentration also shows up in the nearly linear variation of the laterally-averaged concentration fields $\langle \delta C \rangle (z) = \delta \widehat{C}_0(z)$ (dashed line in Fig. 4b) at the plates with a gradient $-\psi N$. The consequence of the plume structure caused by bending the upwards (downwards) concentration jet laterally along the plate can be seen in the SOC state of Fig. 4b by the weak minimum (maximum) of $\delta \widehat{C}_0$ below (above) the boundary layer at the upper (lower) plate. Thus, $\delta \widehat{C}_0$ remains slightly negative (positive) until $z = \frac{1}{2}$. In these regions, the concentration gradient is inverted, giving rise to a weak *destabilizing* effect in the buoyancy force. Inverted concentration gradients have also been seen in thermohaline calculations [74]. Such an inversion is again more common as an effect in the temperature field in high Rayleigh number convection in a pure fluid, e.g., [92]. Note also that slow TW's (third column in Fig. 4b) show a weak inversion of the concentration gradient. For faster TW's (first and second column in Fig. 4b), the large boundary-layer behaviour in the open streamlines of $\widetilde{\mathbf{u}}$ dominates.

The Rayleigh number variation of the Fourier modes at selected vertical positions is displayed in Fig. 5. For a pure fluid, the velocity modes $|\widehat{w}_1|$, $|\widehat{w}_2|$, and $|\widehat{w}_3|$ grow with $r$ [89]. In our binary mixtures, this also holds sufficiently far away from the TW saddle. Approaching the latter from above, in particular $|\widehat{w}_3|$ shows a marked increase (full triangles in Fig. 5a), which is combined with a strong, but smooth phase shift for the TW's, while for the enforced SOC's, $\widehat{w}_3$ goes through zero, combined with a phase jump by $\pi$. There is no significant change of the behavior at the transition from the TW to the SOC branch at $r^*$. In a pure fluid, the temperature modes $|\widehat{\theta}_1|$, $|\widehat{\theta}_2|$, and $|\widehat{\theta}_3|$ also grow with $r$. Note that the temperature is scaled with $\Delta \underline{T}$, so that $|\widehat{\theta}_1|$ decreases with $r$ above a value $r \sim r^*$. At the saddle, there is only a small difference for the states in the mixture compared to pure fluid convection. However, one can see a strong increase in the concentration modes close to the saddle. The reason for this increase lies in the smaller velocity field at the saddle and with that a smaller mixing effect in the concentration field. The structure of the SOC states with or without phase-fixing boundaries is the same on the entire upper branch with a small increase of the concentration modes close to the saddle $r_{SOC}^s$.

The variation of concentration Fourier modes with TW frequency is shown in Fig. 6. In the bulk of the layer, $|\delta \widehat{C}_1|$ grows roughly linearly with $\omega$ (Fig. 6a), which was also observed in experiments [15]. The deviation of $|\delta C_1|$ at $z = 0.25$ from the bulk value at $z = 0.5$ for large $\omega$ reflects the increase of open streamlines of the velocity field when approaching the TW saddle. At the plates (see $z = 0$) we always have boundary-layer behavior, so that the concentration variation with $\omega$ is small. The concentration wave is phase shifted somewhat more than $\frac{\lambda}{4}$ with respect to the velocity wave in the bulk of the layer (Fig. 6b). This is also seen in the experiments of Winkler and Kolodner [15]. For very small frequencies, there is an increase of the phase shift in the bulk to $\frac{\lambda}{2}$ for SOC states. At the plates (see $z = 0$), however, the concentration wave is shifted significantly less than $\frac{\lambda}{4}$ to the velocity wave for TW states and shows no phase shift for SOC states. The laterally-averaged concentration field is nearly zero in the bulk (see Fig. 6c for $z = 0.25$). Thus, the shrinking of the area of closed streamlines and the corresponding smaller mixing area of the concentration has only a weak influence there. This smaller mixing effect can be seen better in the concentration contrast between the plates, which increases linearly with $\omega$ for not too small $\omega$.

### E. Mean flow and time averaged currents of heat and concentration

In this section, we discuss mean transport properties of TW's. In a TW, mass (fluid) is transported along the streamlines of $\widetilde{\mathbf{u}}$ in the comoving frame $\widetilde{\Sigma}$. Thus, in the laboratory system, part of the fluid moves to the right, while another part moves to the left. Here, however, we do not consider this *Lagrangian* motion [8,9,57,56,55,79] but the time-averaged mass transport through an area $S$,

$$\left\langle \int_S d\mathbf{s} \cdot \underline{\rho} \, \underline{\mathbf{u}} \right\rangle = \underline{\rho}_0 \int_S d\mathbf{s} \cdot \langle \underline{\mathbf{u}} \rangle , \qquad (3.29)$$

taking the mass density to be constant. So we discuss the time-averaged velocity field $\langle \mathbf{u} \rangle (x, z)$. In a SOC state, this time average once again yields the velocity field. For a TW, temporal and spatial lateral averages are equivalent, because of (3.1), so that $\langle \mathbf{u} \rangle$ and all other averaged quantities $\langle \delta F \rangle$ depend only on $z$ and no longer on $x$. Because of the incompressibility of the fluid, $\langle w \rangle = 0$. Thus, streamlines of $\langle \mathbf{u} \rangle$ (Fig. 7a) run horizontally. The mean flow [55]

$$U(z) = \langle u(x, z; t) \rangle \qquad (3.30)$$

determining the *time averaged lateral mass current* $\rho_0 U(z)$ [54,57] is opposite to the phase velocity of the TW for all $z$. The vertical average $\overline{U} = \int_0^1 U(z)\, dz$ of this mean flow velocity is very small compared to $v_p$ and to the maximal velocity in the fluid, e.g., $\overline{U} \simeq -0.008$ for a TW at $r = 1.246$, $\psi = -0.25$. As shown by Linz et al. [55], the mean flow $U$ is related to the curvature of the phase, $\varphi_{w_1}(z)$, of the dominant velocity mode. See Ref. [54] for the profile of $\varphi_{w_1}(z)$ and of $U(z)$ [96]. A small curvature of $\varphi_{w_1}$ is also present in linear TW's [66,54,68]. But there the curvature is such that the mean flow is *parallel* to the phase velocity [68]. This is compatible with the $r$-variation of $|U(z = 0.5)|$ shown in Fig. 8a. This extremal value of the mean flow does not only go to zero at $r^*$ but also seems to approach zero in the neighborhood of the saddle $r_{TW}^s$. So we expect from the linear theory [54,68] that the mean flow changes sign near $r_{TW}^s$ and has weakly positive values on part of the lower TW branch and at $r_{osc}$.

The time-averaged heat current,

$$\langle \mathbf{Q} \rangle (z) = N \mathbf{e}_z + \langle \delta T\, u \rangle (z)\, \mathbf{e}_x, \qquad (3.31)$$

is dominated by its vertical contribution $N \mathbf{e}_z$. Also, the mean *convective heat current*

$$\langle \mathbf{Q} - \mathbf{Q}_{cond} \rangle (z) = (N-1) \mathbf{e}_z + \langle \delta T\, u \rangle (z)\, \mathbf{e}_x \qquad (3.32)$$

is dominated by the vertical contribution. But the streamlines of $\langle \mathbf{Q} - \mathbf{Q}_{cond} \rangle$ are bowed in the bulk of the fluid into the direction of the phase velocity of the TW (Fig. 7b): in addition to the *z-independent* vertical part, the TW generates a *z-dependent* mean lateral heat current $\langle \delta T\, u \rangle$ [72,55,54,57], which is parallel to the phase velocity in the lower half of the fluid layer and opposite to the phase velocity in the upper half. Using a lateral Fourier decomposition of the fields, the averaged lateral heat current is [72,55,54]

$$\langle Q_x \rangle (z) = \langle \delta T\, u \rangle$$
$$= \langle \delta T \rangle\, U + \frac{1}{2k} \mathrm{Im} \sum_{n=1}^{\infty} \frac{1}{n} \delta \widehat{T}_n\, \partial_z\, \widehat{w}_n^* . \qquad (3.33)$$

Since $U(z)$ and also the higher lateral harmonics of $w$ and $\delta T$ are small, the main contribution to $\langle \delta T\, u \rangle$ comes from the first lateral harmonic [55,54,57,48], i.e.,

$$\langle \delta T\, u \rangle (z) \simeq \frac{-1}{2k} \left| \delta \widehat{T}_1 \right| (\partial_z |\widehat{w}_1|) \sin(\varphi_{T_1} - \varphi_{w_1}). \qquad (3.34)$$

So the lateral heat current is driven by the phase shift between the temperature and the velocity wave [72,55], which by the way also exists in linear TW's [54,68]. Since for these small frequencies we have $\varphi_{T_1} - \varphi_{w_1} \sim \omega$ [72,55,57], it is evident that the lateral heat current $\langle Q_x \rangle$ is nearly proportional to the TW frequency. This can be seen directly by comparing the maxima of $\langle Q_x \rangle$ shown in Fig. 8b as triangles with the TW frequency in Fig. 8c. The curvature of $\varphi_{T_1}(z)$ is not very important for the heat current, since it is much smaller than $(\varphi_{T_1} - \varphi_{w_1})(z = 0.5)$. As $\varphi_{w_1}(z)$ it is bowed on the upper TW branch opposite to the propagation direction of the TW, but to a stronger degree [54]. However, for linear TW's close to the bifurcation threshold the curvature of $\varphi_{T_1}(z)$ like $\varphi_{w_1}(z)$ is bowed into the propagation direction of the TW's [68].

Similar as for the heat current, the phase shift between the concentration and velocity waves drives an averaged *lateral* concentration current [72,55,54]

$$\langle J_x \rangle (z) = \langle \delta C\, u \rangle$$
$$\simeq \frac{-1}{2k} \left| \delta \widehat{C}_1 \right| (\partial_z |\widehat{w}_1|) \sin(\varphi_{C_1} - \varphi_{w_1}). \qquad (3.35)$$

Like the lateral heat current, it is roughly proportional to the frequency — compare the dots in Fig. 8b which represent the maxima of $\langle J_x \rangle$ with the TW frequency in Fig. 8c. But here the $\omega$ dependence arises from $\left| \delta \widehat{C}_1 \right|$ and not from the phase difference: the concentration wave is in the bulk of the fluid about a quarter wavelength ahead of the $w$-wave for all TW's considered here (cf. sec. III.D). Since concentration is conserved and because of the impermeability of the plates, there is no averaged *vertical* concentration current, i.e.,

$$\langle \mathbf{J} \rangle (z) = \langle \delta C\, u \rangle (z)\, \mathbf{e}_x, \qquad (3.36)$$

and the streamlines of $\langle \mathbf{J} \rangle$ (Fig. 7c) run horizontally. Note that this concentration current has nothing to do with the mean flow $U$. The former is in the upper half of the fluid layer parallel and in the lower antiparallel to the phase velocity [54,57]. So the lateral concentration current runs opposite to the lateral heat current. The variation of $\varphi_{C_1}(z)$ is stronger and more complicated than for the temperature and velocity field since the concentration field shows strong boundary layer behavior at the plates, cf. Fig. 6b and [54, thick line in Fig. 2c]. For a more detailed comparison of linear and nonlinear TW's cf. [68].

The symmetry of the time-averaged quantities

$$\langle \delta F \rangle (z) = \pm \langle \delta F \rangle (1 - z) \qquad (3.37)$$

follows directly from (3.14) with + for $\langle u \rangle, \langle Q_z \rangle$ and − for $\langle Q_x \rangle, \langle J_x \rangle$. Vertically integrating the mean currents, one finds that the mass current, $\rho_0 \overline{U}\, \mathbf{e}_x$, is very weak, the total heat current,

$$\langle \overline{\mathbf{Q}} \rangle = N\, \mathbf{e}_z, \qquad (3.38)$$

is strong, and the total concentration current vanishes:

$$\langle \overline{\mathbf{J}} \rangle = \mathbf{0}. \qquad (3.39)$$

Some questions concerning the significance of these currents in extended convective states are not answered.

The mean flow $U$ is very small – does it play a role? How does the associated mass current look in a finite cell? Does it have an influence there? Does one have to consider spatial density variations [55,57]? This goes beyond the Oberbeck-Boussinesq approximation, which is the basis of our calculations. If one takes these density variations into account in the averaged *lateral* mass current, one should do this also in the *vertical* mass current without violating mass conservation [57]. The heat current is a local current. But what role does the lateral large-scale concentration current play for TW's? Because of concentration conservation, it has to turn around at the lateral cell boundaries. Is the deformation of the streamlines at the boundaries more important? In any case, the concentration current plays an important role in the dynamics of LTW's [46].

## IV. CHANGING THE SORET COUPLING

In sec. III, we discussed a typical bifurcation diagram and the structure of convective states for $\psi = -0.25$. Now we discuss how the bifurcation behavior changes in ethanol-water mixtures with different negative $\psi$. We also show the TW and SOC structure for a small negative $\psi = -0.01$.

For experimentally accessible [3] negative separation ratios $-0.6 \leq \psi \leq -0.005$, we display the bifurcation behavior in the $\psi$-$r$ plane in Fig. 9b. To resolve different orders of magnitude, logarithmic scales are used. For orientation, the bifurcation threshold of TW's, $r_{osc}$, is included. In the $\psi$ region considered here, the conductive state is stable for $r < r_{osc}$. Consider first the behavior for our reference value $\psi = -0.25$. The $r$ region with stable TW's is limited from below by the saddle $r_{TW}^s$ (full dots in Fig. 9b) and from above by the transition to stable SOC states at $r^*$ (full squares in Fig. 9b). The upper SOC branch is stable only above $r^*$. But it can be stabilized down to the SOC saddle $r_{SOC}^s$ (open squares in Fig. 9b) by using phase pinning boundary conditions that prevent TW's. The band limits $r^*, r_{TW}^s$ of stable TW's and $r_{osc}$ grow as $\psi$ becomes more negative. We found

$$0.5 < \frac{r_{TW}^s(\psi) - 1}{r_{osc}(\psi) - 1} < 0.7, \qquad (4.1)$$

i.e., the TW saddle varies with $\psi$ similarly to $r_{osc}(\psi)$. On the other hand, the upper boundary $r^*$ increases strongly at negative $\psi$: $r^*(\psi = -0.25) \simeq 1.65$, while for $\psi = -0.295$ TW's exist up to $r \simeq 5$, and for $\psi = -0.3$ the Soret coupling is so strong that TW's exist there at least up to $r = 10 \approx 7 r_{osc}$. For reasons of numerical accuracy and stability, we did not investigate larger $r$. This very strong increase of $r^*$ suggests the existence of a threshold $\psi$ below which no SOC state is stable. Except for this extreme increase of the region of stable TW's, there is no qualitative difference in the bifurcation structure compared to $\psi = -0.25$.

All quantities $r_{osc}$, $r^*$, and $r_{TW}^s$ decrease with growing $\psi$, and $r^*$ and $r_{osc}$ intersect at $\psi^* \simeq -0.09$. Thus, upon crossing the Hopf bifurcation threshold $r_{osc}$, the initial transient growth of the fast propagating linear TW saturates into a nonlinear slow TW (SOC) state on the upper bifurcation branch for $\psi < \psi^*$ ($\psi > \psi^*$). Fig. 9a shows the frequency of these nonlinear states at $r_{osc}$. The TW frequency at $r_{osc}$ could be fitted with the ansatz $\frac{\omega}{\omega_H}(r_{osc}(\psi)) = a \ln(\psi/\psi^*)$ using $a = 0.25 \pm 3\%$, $\psi^* = 0.091 \pm 0.5\%$. This corrects the wrong formulation in footnote 1 on page 92 of [49]. Tab. I lists besides $r_{osc}$ and $\omega_H$ properties of the nonlinear state at $r_{osc}$. With further increase of $\psi$, $r^*$ approaches the saddle $r_{TW}^s$, so that for $\psi = -0.01$ TW's exist only in a very small interval, $10^{-5} < \Delta r < 10^{-4}$, close to the saddle. Furthermore, the SOC saddle $r_{SOC}^s$ is only just below the TW saddle. But for $\psi = -0.005$ we did not find any stable TW. We did not investigate smaller negative $\psi$. Thus, we conclude that the end point of the TW branch $r^*$ lies on the upper SOC branch for $\psi \leq -0.01$. For negative $\psi \geq -0.005$, the endpoint has moved around the SOC saddle and lies on the lower SOC branch. So we cannot stabilize the SOC's further with phase pinning boundary conditions, and we identify the SOC saddle with the last stable SOC. It should be mentioned that decaying SOC's to the left of the stable SOC saddle eventually end as decaying to the conductive state as *standing waves*.

The structure of TW and SOC states for small Soret coupling, $\psi = -0.01$ (Fig. 10), is in principle the same as for the strong one, $\psi = -0.25$ (Fig. 3). The main difference is that the concentration variation is much smaller, and that there exist stable TW's and SOC's at a much smaller drive $r$. So the strength of convection and of the variation in the temperature field (top row of Fig. 10) is much smaller than for the states with the higher drive (top row of Fig. 3). Similarly, the convective part of the concentration current is not as dominant as for stronger drive. So the concentration boundary layer at the plates and between the rolls is broader (third row of Fig. 10), and the little dips and peaks on the concentration profiles of Fig. 3 have grown here (fourth row of Fig. 10) into broad bumps, and the plume-like structure is much "softer" here than for the case $\psi = -0.25$ with strong convection. Since the TW frequency is very low, the area of open streamlines of $\tilde{\mathbf{u}}$ is small and with it also the concentration difference between the rolls. So the concentration field is governed more by the plume-like structure than by the concentration plateaus. Since the concentration field is weaker and smoother than for $\psi = -0.25$, it contributes less to the sideview shadowgraph intensity (second row of Fig. 10), which hence reflects predominantly the smooth harmonic temperature field.

## V. POSITIVE SORET COUPLING

As mentioned in the introduction, squares and oscillations between perpendicular rolls can be observed for $\psi > 0$. We cannot describe these 3d effects with our 2d algorithm. Our calculations should be compared with measurements in narrow convection channels that favour parallel rolls. Here we show among other things that the structure of the concentration field of SOC states, with not too small amplitudes, can be explained by the same mechanisms for positive $\psi$ that we have discussed for negative $\psi$.

For positive $\psi$, there is only a stationary instability (Fig. 1). The bifurcation threshold $r_{stat}$ becomes small for not too small $\psi$, while $k_c$ drops to zero, so that the wavelength with largest growth rate in a real system is determined by its length. Here we consider $\psi = 0.1$, where $k_c = 0$ and $r_{stat}(k_c) = 0.04216$. The stability curve $r_{stat}(k)$ is very flat, so that the bifurcation of SOC rolls of wavelength $\lambda = 2$ occurs already at $r_{stat}(k = \pi) = 0.06012$ [68]. We have calculated four convective states in a system of length $\Gamma = 20$ with laterally periodic boundary conditions. We have chosen this large $\Gamma$ to allow for wavelength selection. As initial conditions, we took in each case a localized traveling wave at $\psi = -0.25$ [46]. In each case, a SOC state of wavelength $\lambda = 2$ finally developed (squares in Fig. 11), thus suggesting the existence of a strong selection mechanism. Only the state with the biggest $r$-value was calculated in a system of length $\Gamma = 2$. The first two states in Fig. 11 are in the so called Soret region, $r < 1$, and the others in the Rayleigh regime, $r > 1$. Note that even the smallest $r$-value (r=0.506) lies distinctly above the bifurcation threshold. As expected [26], the convective heat current $N - 1$ (Fig. 11b) in the Soret region is very small, while in the Rayleigh region it is comparable to the current in a pure fluid. The mixing parameter $M$ (Fig. 11a) indicates that the fluid in the Rayleigh regime is significantly better mixed than in the Soret regime. But $M$ being well below 1 clearly shows that the states in the Soret regime also have to be considered as "strongly nonlinear states".

Now we discuss structural variations of SOC states with $r$. Consider first the state in Fig. 12d at $r = 2.026$ with the largest amplitude. Its concentration is well mixed in the rolls as in the SOC states at negative $\psi$ (cf. Fig. 3d). Also, the concentration boundary layer is present, as at negative $\psi$. But here $\delta C$ is positive at the lower plate and negative at the upper plate, since $\psi > 0$. Also, the little peaks in the lateral concentration profile (fourth row of Fig. 12d) now lie at the position of maximal *up*flow and symmetrically the valleys at the positions of maximal *down*flow.

With decreasing $r$ (see Fig. 12c, $r = 1.114$), the velocity field decreases, the concentration boundary layer grows, and the little peaks become higher and broader. But mixing in the rolls remains dominant down to the beginning of the Soret region (see Fig. 12b, $r = 0.912$). However, here the plume structure is very soft, as in the SOC state at $\psi = -0.01$ in Fig. 10b, since the velocity field is very weak.

With further decrease of $r$ and thereby of the velocity field, concentration diffusion becomes comparable to advection in a larger spatial area, the concentration plumes become very broad (Fig. 12a, $r = 0.506$), and the lateral concentration profile (third row of Fig. 12a) becomes smooth. Since convective mixing is weak, the amplitude of the lateral concentration profile at $z = \frac{1}{2}$ has increased, and the vertical concentration difference between the plates has also increased to almost half the size in the quiescent conductive state. With further decrease of $r$, the nonlinear advective term in the concentration balance loses more and more dominance and then relevance, and the mixing parameter $M$ increases towards 1 at the bifurcation threshold.

The second and last rows of Fig. 12 show the shadowgraph intensity $I(x, z)$ (3.15) with $b = -0.919$. The temperature field dominates in the Rayleigh regime, and the concentration field adds sharp features to it, as for not too weakly negative $\psi$ (Fig. 3d and ref. [14]). But here we don't have a single maximum at $x = \pm 1$, but two light spots, i.e., two symmetric maxima around the positions of maximal downflow. They are caused by the concentration field. Its contribution to the shadowgraph intensity is opposite in sign to the temperature contribution at the downflow roll boundaries for $\psi > 0$, since $b$ is negative [3] for all ethanol-water mixtures considered so far — for simplicity and better comparison we use here the same $b$ as for the mixture with $\psi = -0.25$. In SOC's with $\psi < 0$, both contributions are positive, giving rise to the single maximum in Figs. 3d, 10b.

Comparing a shadowgraph image in the Rayleigh regime (Fig. 12d) with one in the Soret regime (Fig. 12a), one sees that bright areas have become dark, and vice versa. This again reflects the facts that (i) concentration and temperature fields contribute basically with opposite signs to $I(x, z)$, and that (ii) the former field is stronger than the latter field in the Soret regime, and vice versa high up in the Rayleigh regime. But the two bright spots for the smallest $r$-value in Fig. 12a at $x = \pm 1, z \approx 0.75$ indicate that concentration as well as temperature field contribute to the shadowgraph intensity, and that the concentration field is still strongly anharmonic. See also the lateral concentration profile in the fourth row of Fig. 12a.

We should mention that in our calculations we did not wait for perfect stationarity of the SOC states. For the three states with the higher $r$, we observed a very slow relaxation process in the concentration field with a sinusoidal spatial modulation given by the system length $\Gamma = 20$, while changes in the temperature and velocity field were nearly negligible. To filter out these long-time variations of $\delta C$, we presented in Fig. 12c an average of $\delta C$ over 10 wavelengths. Further work is necessary to investigate whether this relaxation process reflects slow phase diffusion or weakly damped concentration modes.

A reason for using localized *traveling* waves as initial conditions was to test if there exist TW's in the Rayleigh region for positive $\psi$ as predicted by the Galerkin model of Linz et al. [55] for free-slip impermeable boundary conditions. But in a more systematic search in the Rayleigh region for fluids with $\psi = 0.02$ and $\psi = 0.05$, we also did not find stable TW's. Furthermore, test simulations with free-slip impermeable boundary conditions did not show stable TW's for positive $\psi$. On the other hand, Bigazzi et al. [22] found oscillatory behavior in their cylindrical convection cell which resembled TW's. Since oscillatory behavior in a cylindrical cell may have different causes, we think there is now more evidence against stable TW's for positive $\psi$. Experiments in annular cells are most desirable to clarify this issue.

We did not perform calculations right above the instability despite the prospect of interesting wave number selection processes with possibly hysteretic character, since for our long system calculations are very time consuming, and since 3d effects have been predicted close to the instability [76,77,62].

## VI. LEWIS- AND PRANDTL NUMBER DEPENDENCE

In the preceding sections, we have discussed convective properties of binary mixtures as functions of the control parameters $r$ and $\psi$, while keeping the Lewis number $L = 0.01$ and the Prandtl number $\sigma = 10$ fixed. Yet $L$ varies between 0.005 and 0.01 for ethanol-water mixtures in the temperature range between $10^\circ C$ and $40^\circ C$ [3], while $\sigma$ varies between 5 and 20 [3]. On the other hand, $L = 0.03$ and $\sigma = 0.6$ in $^3$He-$^4$He-mixtures, where experiments have also been performed. Furthermore, Steinberg et al. [30] report experiments with a benzene-methanol mixture at about $L = 0.025$, $\sigma = 7.5$, $\psi = -0.045$.

We start the investigation of the $L, \sigma, \psi, r$ parameter space with the bifurcation diagrams of Fig. 13, each of which is for $\psi = -0.25$. Characteristic properties for the four $L - \sigma$ combinations discussed in the following are given in Tab. II. Let us start from $L = 0.01$, $\sigma = 10$ (Fig. 2 and right curves of Fig. 13a,b) and increase the Lewis number to the $^3$He-$^4$He value $L = 0.03$, but keeping $\sigma = 10$ constant (right curves of Fig. 13c,d). The Hopf frequency $\omega_H$ and stability boundary $r_{osc}$ [68] increase only slightly despite of the tripling of $L$. The saddle $r_{TW}^s$ increases slightly, and the frequency at the saddle stays roughly constant. But $r^*$ decreases strongly, and thereby roughly halves the region $r^* - r_{TW}^s$ of stable TW's on the upper branch. But $r^*$ still lies slightly above $r_{osc}$, so that the nonlinear TW at $r_{osc}$ has a small, but finite frequency $\omega(r_{osc}) \approx \frac{1}{30}\omega_H$.

Let us now go back to our ethanol-water reference fluid ($L = 0.01$, $\sigma = 10$) and reduce the Prandtl number to the $^3$He-$^4$He value $\sigma = 0.6$, while keeping $L = 0.01$ constant. For these parameters, $r_{osc} - 1$ nearly halves, and $\omega_H$ decreases to about $\frac{2}{3}$ of the ethanol-water value (left curves of Fig. 13a,b). The saddle lies closer to $r_{osc}$, so that the hysteretic TW region $r_{osc} - r_{TW}^s$ has become smaller. The frequency at the saddle has decreased to about $\frac{1}{4}\omega_H$, compared to about $\frac{1}{3}\omega_H$ for the ethanol-water mixture. Since $r^*$ lies just above $r_{osc}$, the region of stable TW's is significantly smaller than for ethanol-water with the higher Prandtl number.

The two effects of (i) increasing $L$ and (ii) reducing $\sigma$ all in all "add up" for the $^3$He-$^4$He parameters $L = 0.03$, $\sigma = 0.6$ (left curves in Fig. 13c,d). Note here that $r^*$ lies *below* $r_{osc}$, i.e., the final state is nonlinear SOC. The region, $r^* - r_{TW}^s$, of stable TW's on the upper branch is smaller by a factor of 10 than for ethanol-water parameters, also if we scale with $r_{osc} - 1$. But the qualitative structure of the bifurcation diagrams is the same for the four parameter combinations of Fig. 13. In particular there exists an upper TW branch.

A more global picture of the changes of various bifurcation properties with varying $L$ and $\sigma$ is shown in the $L - r$ plane of Fig. 14b and in the $\sigma - r$ plane of Fig. 15b. See also tables III and IV. With increasing $L$, the TW bifurcation threshold $r_{osc}$ first increases slightly and then strongly at $L = O(10^{-1})$ (Fig. 14b). The Hopf frequency $\omega_H(k = \pi)$, on the other hand, remains essentially constant for $0.001 < L < 0.1$. The saddle $r_{TW}^s$ lies roughly parallel to $r_{osc}$ in Fig. 14b, while $r^*$ decreases very strongly with increasing $L$. Thus, at $L = 0.05$, the transition to stationary convection is already below $r_{osc}$ (see also the frequency of the nonlinear state at $r_{osc}$ in Fig. 14a), and for $L = 0.3$ there is no longer a stable upper TW branch, although $r^*$ has increased with increasing $r_{osc}$. The vanishing of the upper TW branch here is similar to the vanishing in Fig. 9 for $\psi = -0.005$. Note the rapid expansion of the region of stable TW's with decreasing $L$ in Fig. 14b: the upper boundary $r^*$ shoots up from 1.65 at $L = 0.01$ to more than 5 at $L = 0.005$. This resembles the rapid increase of $r^*$ with decreasing $\psi$ around $\psi = -0.3$ shown in Fig. 9b. The above described rapid variation with $L$ presumably accounts for part of the differences that Liu and deBruyn [16] observed between their experiments at $L = 0.005$ and our calculations at $L = 0.01$. To understand the increase of the TW region with decreasing $L$, the boundary layer model of Bensimon et al. [51] might be a good starting point. However, in its present form, it fails to give the ending of the TW branch on the *lower* SOC branch and the extreme increase of $r^*$, as we have seen in the $L$ and $\psi$ variation.

Let us now consider the variation with $\sigma$ while we keep $L = 0.01$ constant. With decreasing $\sigma$, the Hopf frequency and $r_{osc}$ decrease, but below $\sigma \approx 0.14$, the bifurcation threshold $r_{osc}$ increases again. Also, the frequency of the nonlinear state at $r_{osc}$ decreases (Fig. 15a); indeed, it decreases faster than the Hopf frequency (see Tab. IV). In the considered parameter region, $r^*$ decreases down to the order of 1 with falling $\sigma$ like $r^* \approx a + b \log(\sigma)$. Down to $\sigma = 0.5$, there exist still stable TW's above $r_{osc}$, and stable TW's close to the saddle exist at least down to $\sigma = 0.1$. Note that, in this small Prandtl-number regime, strong nonlinearities can also be seen in the *velocity* fields of TW and SOC states, which deserve further investigation.

Let us now address $^3$He-$^4$He mixtures with $L = 0.03$ and $\sigma = 0.6$ [36–45]. We have investigated their dependence on $\psi$ only in the range $-0.25 \leq \psi \leq -0.05$. As we have discussed above, the qualitative structure of the bifurcation diagrams for $\psi = -0.25$ is the same as for ethanol-water mixtures, but here $r^*$ lies just *below* $r_{osc}$. This reduction of the region of stable TW's can also be seen in the variation with $\psi$: while we still see stable TW's for $\psi = -0.1$, there are no longer stable TW's for $\psi = -0.05$. In comparison, in ethanol-water mixtures, there are stable TW's even for $\psi = -0.01$.

## VII. CONCLUSION

The main goal of this paper has been to provide a quantitative description of extended convection in binary fluid mixtures and thus come to an understanding of the nonlinear stationary and traveling-wave states that occur as primary convective states near the convective threshold. In particular, we wanted to elucidate the role of the concentration field. We should like to stress that the experimentally observable convective states on the upper stable parts of the bifurcation branches that we have investigated are strongly nonlinear. For example, for most of them, the concentration Péclet numbers $\underline{w}d/D = w/L$ measuring the magnitude of the nonlinear advective contribution to the concentration balance equation relative to the linear diffusive one were of order 1000. The field structure of these states can be described quantitatively neither by a weakly-nonlinear expansion around their bifurcation threshold nor by simple, heuristic, one-mode amplitude-equation approaches. We therefore solved the full hydrodynamic field equations numerically with realistic boundary conditions, using an explicit finite-differences method. We restricted ourselves to 2d convection in the form of straight rolls by suppressing spatial variations along the roll axes. In most cases, we enforced the wavelength of the states, by imposing laterally periodic boundary conditions, to be $\lambda = 2$, i.e., close to the wavelengths seen in experiments.

*Typical bifurcation behaviour for negative Soret-coupling* — In a bifurcation diagram of convective intensity versus reduced Rayleigh number, an unstable TW solution branch bifurcates backwards at a Hopf bifurcation threshold $r_{osc}$ out of the conductive state. This TW branch turns over at a saddle $r^s_{TW}$ below $r_{osc}$ and becomes stable. The upper stable branch ends by merging at $r^*$ above $r^s_{TW}$ into the SOC solution branch. Moving along the TW branch, the TW frequency continuously decreases from the Hopf frequency at threshold down to zero at the end point $r^*$. The SOC branch bifurcates at $r_{stat}$ backwards out of the conductive state or is disconnected from it after $r_{stat}$ has reached infinity. This lower branch bends forwards at a saddle $r^s_{SOC}$. By suppressing TW convection with phase pinning boundary conditions, we determined the upper SOC branch all the way down to its saddle. However, allowing for phase propagation, SOC's are unstable below $r^*$: for $r < r^s_{TW}$ they start propagating and decay into the conductive state, and in the interval $r^s_{TW} < r < r^*$, they start propagating and end up as stable TW's. Viewed from the SOC branch above $r^*$, this branch loses stability in a mirror-symmetry-breaking bifurcation to a stable left or right TW whose frequency grows $\sim (r^* - r)^{\frac{1}{2}}$ close to the bifurcation threshold $r^*$, in good agreement with experiments in ethanol-water mixtures in an annular cell [5,11]. These annular cells are closed convection channels with a rectangular cross-section. Apart for the channel walls, there are no lateral walls which hinder the traveling of the TW's as in laterally finite channels. The transition TW $\leftrightarrow$ SOC at $r^*$ is not forced by lateral walls but is an intrinsic effect which was described by a concentration boundary layer expansion [51] around the states of steady convection in a pure fluid. This model of Bensimon et al. [51] can describe this transition and shows the connection of the upper TW branch via the lower TW branch with the conductive state at $r_{osc}$.

*Structure of the fields* — Both the velocity and temperature fields of SOC's and of TW's look similar in structure and in strength to those in a comparable stationary state in pure-fluid convection. But in the TW state of the mixture, the temperature wave is slightly phase-shifted relative to the velocity wave. The shift varies nearly linearly with the TW frequency as predicted earlier [72,55]. The lines of constant phase of the first lateral TW Fourier modes being curved, the waves of temperature, concentration and velocity are not planar.

While velocity and temperature waves are nearly harmonic, the concentration wave has a strongly anharmonic, trapezoidal profile. The latter shows that, in a TW, the concentration is almost homogeneous within each roll, but at alternatingly high and low concentration levels, with a linear variation between adjacent rolls. This structure can best be understood within the reference frame that comoves with the TW's phase velocity $v_p$, so that all fields are stationary therein. With the diffusion constant being small, concentration is transported predominantly passively by the velocity field. Thus, one can qualitatively infer from its streamlines in the comoving frame the concentration distribution in a TW. Because of the contribution $-v_p z$ to the stream function from the phase velocity, the roll-like regions with closed streamlines are shifted alternatingly towards the top and bottom plates, and in addition there are open streamlines. They meander between the roll-like regions and separate them from each other and from the plates, in such a way that the boundary layer with the Soret-induced concentration surplus (deficiency) at the top (bottom) plate feeds, for instance, in a right TW only the right (left) turning roll regions: high (low) concentration plumes emanating from the top (bottom) boundary layer are bent into the right (left) turning roll regions where the fluid is mixed up. So the concentration wave profile resulting from this combination of boundary layer behaviour perpendicular to the open velocity streamlines and mixing behaviour within the regions of closed streamlines cannot be described by a few lateral Fourier modes, $e^{-inkx}$, as the velocity or temperature wave. This clearly shows that stable TW's on the upper bifurcation branch are strongly nonlinear states whose structural dynamics differs significantly from marginally-stable linear TW modes at the Hopf bifurcation threshold.

The phase velocity or frequency of the TW is the single most relevant parameter that governs the structural properties of TW's. With decreasing $v_p$, the streamlines of the velocity field in the comoving frame become more and more mirror symmetric around up- and downflow positions, the regions of closed (open) streamlines grow (shrink), their alternating up- and downwards shift from the middle of the layer is reduced, the feeding of concentration into the different roll regions becomes more symmetric, and the concentration contrast between adjacent rolls — and with it the plateau-height in the trapezoidal profile of the concentration wave — is reduced until at $r^*$ the SOC concentration distribution is perfectly mirror symmetric around up- and downflow. Then concentration plumes ascending (descending) at the positions of maximal up (down) flow from the bottom (top) boundary layer inject low (high) concentration symmetrically into left- and right-turning rolls, where the fluid is well mixed in the bulk of the layer. The symmetric mixing operation of SOC's homogenizes the concentration more efficiently than the asymmetric mixing at different levels in adjacent TW rolls. This can be quantified by our mixing parameter $M$ measuring the rms variations of the concentration field.

The plume-like transport of concentration out of the boundary layers near the plates into the bulk of TW's and SOC's is a consequence of the smallness of the diffusion constant and thus is similar to thermal plumes in high Rayleigh number convection, where convective heat transport dominates in the bulk over diffusive transport.

*Symmetry properties* — TW's and SOC's show a mirror glide symmetry under lateral translation by half a wave-length combined with reflection through the horizontal midplane of the layer. This can easily be seen, e.g., in side-view shadowgraphs [15] which, by the way, yield detailed information on the concentration distribution in the fluid.

*Mean transport of TW's* — TW's cause a lateral mean flow $U(z)$ that is several orders of magnitude smaller than $v_p$ and that is related [55,54] to the $z$-variation of the phase of the first lateral Fourier mode of the lateral velocity field. On the other hand, and more importantly, the phase shift between vertical velocity and temperature (concentration) wave generates a significant mean lateral convective heat current $\langle u\,\delta T\rangle$ (concentration current $\langle u\,\delta C\rangle$) that flows parallel (opposite) to the TW propagation in the upper half of the fluid layer and vice versa in the lower half, so that the net currents vanish. Both current amplitudes roughly grow proportional to $v_p$. The lateral convective transport of heat and concentration by these currents is different from and unrelated to the mean flow $U(z)$.

*Changing the Soret coupling* — The upper (lower) existence limit $r^*$ ($r^s_{TW}$) of TW's increases strongly (slightly) as the coupling parameter $\psi$ becomes more negative. On the other hand, as the magnitude of the negative separation ratio $\psi$ is reduced, the TW end point $r^*$ moves downwards on the SOC branch, so that $r^*$ approaches the saddles $r^s_{SOC}$ and $r^s_{TW}$. The upper part of the TW branch shrinks to zero when $r^*$ and $r^s_{TW}$ meet. That happens close to $r^s_{SOC}$. In ethanol-water mixtures ($L=0.01, \sigma=10$), we found for $\psi=-0.01$ still a small $r$-range of stable TW's, cf. [49, Fig. 5.4], but not for $\psi=-0.005$. So we conclude that for $\psi \geq -0.005$ the TW endpoint has already moved around the SOC saddle and lies on the lower SOC branch, while the upper SOC branch has become stable all the way down to $r^s_{SOC}$. A negative Soret coupling that is smaller in magnitude implies smaller concentration variations with smoother profiles, broader boundary layers, and wider plumes.

*Positive Soret coupling* — Calculations done for $\psi=0.1$ in a system of lateral periodicity length $\Gamma=20$ showed in the Soret regime, $r<1$, but well above the instability and also in the Rayleigh regime, $r>1$, always an evolution into SOC states with wavelength $\lambda=2$. This should be compared with experiments in narrow convection channels that suppress squares and favour rolls. We used localized traveling waves as initial conditions to test if there exist stable TW's at $\psi=0.1$ but found none — also not with free-slip impermeable boundary conditions. The characteristic structure of the SOC concentration field — boundary layers, plumes, and mixing within the rolls — appears as for negative Soret coupling. Here, however, there is a concentration surplus at the lower, warm plate. Reducing $r$ into the Soret regime, the flow intensity decreases, diffusive transport contributes more and more, concentration profiles become smooth, and boundary layers and plume structures become broad. But SOC's in the Soret regime are still nonlinear states as documented by the mixing parameter $M$ being well below the conductive reference value $M_{cond}=1$.

*Lewis- and Prandtl number dependence* — Our calculations at $\psi=-0.25$ done for $0.005 \leq L \leq 0.3$ and $0.1 \leq \sigma \leq 10$ showed that, upon decreasing $L$ or increasing $\sigma$ beyond the reference values $L=0.01, \sigma=10$ of ethanol water, the existence range of stable TW's on the upper branch widens with increasing $r^*$. On the other hand, increasing $L$ or decreasing $\sigma$ eventually pinches off their existence range as $r^*$ and $r^s_{TW}$ come together. But otherwise the bifurcation structure of TW's and SOC's remains unchanged for negative Soret coupling.

## ACKNOWLEDGMENTS

Discussions with G. Ahlers, K. D. Eaton, S. Hollinger, W. Hort, P. Kolodner, S. J. Linz, H. W. Müller, D. R. Ohlsen, D. Roth, F. K. Schank, H. Schmitt, and C. M. Surko are gratefully acknowledged. We thank P. Kolodner for carefully reading the manuscript and for suggesting improvements. This work was supported by the Deutsche Forschungsgemeinschaft and the Stiftung Volkswagenwerk.

TABLE I. $\psi$-dependence in ethanol-water mixtures ($L = 0.01$, $\sigma = 10$). Given are $r_{osc}(k=\pi)$ and $\omega_H(k=\pi)$ [68]. The last four columns contain properties of the nonlinear convective states on the upper bifurcation branches at $r_{osc}$.

| $\psi$ | $r_{osc}$ | $\omega_H$ | $N-1$ | $w_{\max}$ | $\omega$ | $\frac{\omega}{\omega_H}$ |
|---|---|---|---|---|---|---|
| -0.6 | 2.4120 | 23.4391 | 0.8039 | 14.290 | 8.6852 | 0.3705 |
| -0.3 | 1.4258 | 12.7229 | 0.4362 | 7.839 | 1.6525 | 0.1299 |
| -0.25 | 1.3347 | 11.2351 | 0.3639 | 6.896 | 1.2240 | 0.1089 |
| -0.2 | 1.2528 | 9.7405 | 0.2919 | 5.972 | 0.8526 | 0.0875 |
| -0.1 | 1.1201 | 6.5031 | 0.1495 | 4.019 | 0.0696 | 0.0107 |
| -0.08 | 1.0966 | 5.7534 | 0.1211 | 3.580 | 0 | 0 |
| -0.05 | 1.0630 | 4.4746 | 0.0790 | 2.850 | 0 | 0 |
| -0.03 | 1.0418 | 3.4262 | 0.0515 | 2.278 | 0 | 0 |
| -0.02 | 1.0314 | 2.7787 | 0.0382 | 1.956 | 0 | 0 |
| -0.01 | 1.0212 | 1.9451 | 0.0256 | 1.595 | 0 | 0 |
| -0.005 | 1.0159 | 1.3587 | 0.0198 | 1.402 | 0 | 0 |

TABLE II. Linear and nonlinear properties of the bifurcation diagrams in Fig. 13 for $\psi = -0.25$ are given for four combinations of $L$ and $\sigma$.

| $L$ | 0.03 | 0.03 | 0.01 | 0.01 |
|---|---|---|---|---|
| $\sigma$ | 10 | 0.6 | 0.6 | 10 |
| $r_{osc}(k=\pi)$ | 1.3651 | 1.2315 | 1.1833 | 1.3347 |
| $r_{TW}^s$ | 1.228 | 1.159 | 1.127 | 1.213 |
| $r^*$ | 1.375 | 1.185 | 1.198 | 1.65 |
| $\omega_H(k=\pi)$ | 11.3207 | 8.0649 | 7.9512 | 11.2351 |
| $\omega(r_{TW}^s)$ | $\approx 3.8$ | $\approx 2.5$ | $\approx 2.0$ | $\approx 3.8$ |
| $\frac{\omega(r_{TW}^s)}{\omega_H}$ | $\approx 0.33$ | $\approx 0.31$ | $\approx 0.25$ | $\approx 0.34$ |
| $\omega(r_{osc})$ | 0.404 | 0 | 0.413 | 1.224 |
| $\frac{\omega(r_{osc})}{\omega_H}$ | 0.0357 | 0 | 0.0520 | 0.109 |
| $\frac{r_{TW}^s - 1}{r_{osc} - 1}$ | 0.624 | 0.687 | 0.693 | 0.636 |
| $\frac{r^* - 1}{r_{osc} - 1}$ | 1.027 | 0.799 | 1.08 | 1.94 |
| $\frac{r^* - r_{TW}^s}{r_{osc} - 1}$ | 0.403 | 0.112 | 0.387 | 1.30 |
| $w_{\max}(r^*)$ | 7.13 | 4.37 | 5.04 | 9.9 |

TABLE III. $L$-dependence in mixtures with $\sigma = 10$, $\psi = -0.25$. Given are $r_{osc}(k=\pi)$ and $\omega_H(k=\pi)$ [68]. The last four columns contain properties of the nonlinear convective states on the upper bifurcation branches at $r_{osc}$.

| $L$ | $r_{osc}$ | $\omega_H$ | $N-1$ | $w_{\max}$ | $\omega$ | $\frac{\omega}{\omega_H}$ |
|---|---|---|---|---|---|---|
| 0.3 | 1.6328 | 10.7391 | | | | |
| 0.2 | 1.5627 | 11.1155 | | | | |
| 0.1 | 1.4615 | 11.3909 | 0.4309 | 7.513 | 0 | 0 |
| 0.05 | 1.3945 | 11.3747 | 0.3973 | 7.162 | 0 | 0 |
| 0.03 | 1.3651 | 11.3207 | 0.3822 | 7.024 | 0.4045 | 0.0357 |
| 0.02 | 1.3500 | 11.2820 | 0.3725 | 6.961 | 1.0135 | 0.0898 |
| 0.01 | 1.3347 | 11.2351 | 0.3639 | 6.896 | 1.2240 | 0.1089 |
| 0.005 | 1.3271 | 11.2086 | 0.3602 | 6.863 | 1.2566 | 0.1121 |
| 0.003 | 1.3240 | 11.1972 | 0.3585 | 6.864 | 1.2912 | 0.1153 |
| 0.002 | 1.3225 | 11.1915 | 0.3574 | 6.854 | 1.3465 | 0.1203 |
| 0.001 | 1.3209 | 11.1858 | 0.3553 | 6.830 | 1.4960 | 0.1337 |

TABLE IV. $\sigma$-dependence in mixtures with $L = 0.01$, $\psi = -0.25$. Given are $r_{osc}(k=\pi)$ and $\omega_H(k=\pi)$ [68]. The last four columns contain properties of the nonlinear convective states on the upper bifurcation branches at $r_{osc}$.

| $\sigma$ | $r_{osc}$ | $\omega_H$ | $N-1$ | $w_{\max}$ | $\omega$ | $\frac{\omega}{\omega_H}$ |
|---|---|---|---|---|---|---|
| 10 | 1.3347 | 11.2351 | 0.3639 | 6.896 | 1.2240 | 0.1089 |
| 5 | 1.3158 | 10.8911 | 0.3474 | 6.687 | 1.1222 | 0.1030 |
| 3 | 1.2940 | 10.4785 | 0.3278 | 6.436 | 1.0242 | 0.0977 |
| 2 | 1.2711 | 10.0239 | 0.3061 | 6.151 | 0.8979 | 0.0896 |
| 1 | 1.2222 | 8.9526 | 0.2558 | 5.474 | 0.6409 | 0.0716 |
| 0.6 | 1.1833 | 7.9512 | 0.2103 | 4.836 | 0.4134 | 0.0520 |
| 0.5 | 1.1701 | 7.5643 | 0.1933 | 4.580 | 0.3173 | 0.0419 |
| 0.3 | 1.1391 | 6.4491 | 0.1465 | 3.844 | 0 | 0 |
| 0.2 | 1.1237 | 5.5820 | 0.1132 | 3.28 | 0 | 0 |
| 0.1 | 1.1243 | 4.2549 | 0.0745 | 2.516 | 0 | 0 |

FIG. 1. $\psi$-dependence of linear stability properties of the heat conducting state of ethanol-water mixtures [68]. (a) Critical reduced Rayleigh numbers $r_{osc}$ (dashed line) of the oscillatory and $r_{stat}$ (full and dash-dotted line) of the stationary instabilities. (b) Critical reduced wave numbers of the oscillatory (dashed line) and stationary (full line) instability for $r > 0$. The critical wave number of the stationary instability at $r < 0$ vanishes. (c) Critical Hopf frequency $\omega_H$.

FIG. 2. Bifurcation diagrams of extended convective states with wavelength $\lambda = 2$. (a) Frequency, (b) mixing parameter $M$, (c) Nusselt number. The TW solution branch bifurcates backwards (this unstable branch is not shown) at $r_{osc} = 1.3347$ with frequency $\omega_H = 11.235$ [68], becomes stable (full circles) at $r_{TW}^s = 1.213$, and ends at $r^* = 1.65$. SOC's (open squares) are stable only above $r^*$. By phase pinning, they can be stabilized (open triangles) down to $r_{SOC}^s = 1.08$. The lower unstable SOC solution branch (not shown) is disconnected from the conductive solution and lies almost parallel to the $r$-axis. The dotted line in (c) shows the Nusselt number in a pure fluid with $\sigma = 10$.

FIG. 3. Snapshots of three representative TW's traveling to the right and a SOC state for $L = 0.01$, $\sigma = 10$, $\psi = -0.25$. (a) TW at $r = 1.2155$ close to the saddle with $\omega \approx 0.3\omega_H$, (b) TW at $r = 1.2459$ with $\omega \approx \frac{1}{5}\omega_H$, (c) TW at $r = 1.4181$ with $\omega \approx 0.07\omega_H$, and (d) SOC state at $r = 1.8232$. See Fig. 2 to locate the $r$-values in the bifurcation diagrams. The wavelength is $\lambda = 2$, and $x = 0$ denotes the position of maximal upflow. The colour coded temperature field $\delta T$ (first row), with arrows denoting the velocity field **u**, is shown only for the fastest TW (a) and the SOC state (d). The second and seventh row show the sideview shadowgraph intensity $I(x, z)$ (3.15) with $b = -0.919$ in grey scales. The concentration field $\delta C$ in the third and sixth row is given in the same colour code for all four states. Red, green, and blue represent low, mean, and high concentration, respectively. The structure of the concentration field can be related to the dotted streamlines of the velocity field $\tilde{\mathbf{u}}$ in the frame comoving with the phase velocity of the states. Fourth and fifth rows show lateral profiles of $w$ (thin line), $40\,\delta T$ (triangles), and $400\,\delta C$ (squares) at midheight, $z = 0.5$.

FIG. 4. Vertical structure of TW and SOC states. Columns show from left to right the states of Fig. 3: (1) TW at $r = 1.2155$, (2) TW at $r = 1.2459$, (3) TW at $r = 1.4181$, (4) SOC state at $r = 1.8232$. Row (a) shows $z$-dependence of lateral Fourier modes $\widehat{\theta}_0$ and $|\widehat{\theta}_n|$ ($n = 1, 2, 3$) of the temperature field. In row (b), the corresponding modes of the concentration field, $\delta C$, are presented. For better resolution especially at the plates, Fourier modes were evaluated also vertically halfway between grid positions. Row (c) shows vertical profiles of the concentration field $\delta C$ at $x$-positions A = 0.35, B = 0.85 in the right turning roll and A' = $-0.65$, B' = $-0.15$ in the left turning roll differing by $\frac{\lambda}{2}$. Maximal upflow is at $x = 0$ as in Fig. 3. For comparison, the conductive profile is shown as a dotted line.

FIG. 5. Rayleigh-number dependence of lateral Fourier modes $|\widehat{f}_n|$ of: (a) vertical velocity $f = w$, (b) temperature $f = \theta$, (c) concentration $f = \delta C$ for TW (closed symbols) and SOC states (open symbols). SOC's below $r^*$ are stabilized by phase-pinning boundary conditions. Presented are $|\widehat{f}_1|$ at $z = 0.5$ (dots for TW's and squares for SOC's), $|\widehat{f}_2|$ at $z = 0.25$ (lozenges), and $|\widehat{f}_3|$ at $z = 0.5$ (triangles). For comparison, the corresponding modes of SOC states in a pure fluid, $\psi = 0$, are included in (a) and (b) by lines: full for $n = 1$, dotted for $n = 2$, and dashed for $n = 3$.

FIG. 6. Frequency dependence of lateral Fourier modes of the concentration field of TW's at different $z$ positions. (a) $|\delta\widehat{C}_1|$, (b) phase shift between concentration wave and wave of vertical velocity, (c) lateral average $\delta\widehat{C}_0$. In (c), $\delta\widehat{C}_0(z = 0.5) = 0$ is not included. The highest frequency denotes the TW nearest to the saddle, $r_{TW}^s$, while $\omega = 0$ at $r^*$. For $\omega \to 0$, the phase shift in (b) approaches $\pi$ in the bulk ($z = 0.5$ and $z = 0.25$) but drops to zero at the plates ($z = 0$).

FIG. 7. Streamlines in the x-z plane for time-averaged transport in the right-going TW of Fig. 3a. (a) mean flow $\langle \mathbf{u} \rangle$, (b) heat current $\langle \mathbf{Q} \rangle - \mathbf{Q}_{cond}$, and (c) concentration current $\langle \mathbf{J} \rangle$.

FIG. 8. Lateral transport of TW's. (a) Amplitude of lateral flow $|U(z = 1/2)|$. (b) Maxima of the mean lateral currents of heat, $\langle Q_x \rangle$ (triangles), and of concentration, $\langle J_x \rangle$ (circles). (c) Shows the TW frequency $\omega$ for comparison.

FIG. 9. (a) Frequency of nonlinear convective states on the upper bifurcation branch at $r_{osc}$ reduced by the Hopf frequency $\omega_H$. TW's are shown by full dots and SOC states with $\omega = 0$ as full squares. See also Tab. I. (b) Double-logarithmic phase diagram. Symbols come from simulations done with prescribed wavelength, $\lambda = 2$. The conductive state is stable below and unstable above the TW bifurcation threshold $r_{osc}(k = \pi)$ (thin line) [68]. Stable TW's on the upper branch exist in the shaded area between the TW saddle $r_{TW}^s$ (full dots) and $r^*$ (full squares), where the TW branch merges into the SOC branch. SOC states are stable above $r^*$ or — for $\psi$ values close to zero where we have not found stable TW's — above the SOC saddle $r_{SOC}^s$ (line with filled squares in the lower right corner). Where stable TW states are found, SOC states are unstable below $r^*$. But the upper SOC branch can be stabilized down to the SOC saddle $r_{SOC}^s$ (open squares) by suppressing TW's. At $\psi = -0.01$, where $r_{SOC}^s \approx r_{TW}^s \approx r^*$, there still exists a very small area of stable TW's, while for $\psi = -0.005$ we have not found any stable TW.

FIG. 10. Snapshots of convective states at $\psi = -0.01$, $L = 0.01$, $\sigma = 10$. (a) TW at $r = 1.01483$ with $\omega \approx 0.074$, (b) SOC state at $r = 1.0212$. The presentation is as in Fig. 3. However, to resolve the concentration variation, its colour code is different as indicated. The lateral profiles at $z = 0.5$ in the last row show $w$ (thin line), $20\,\delta T$ (triangles), and $3000\,\delta C$ (squares).

FIG. 11. Bifurcation diagrams for positive Soret coupling. The simulations (squares) of a system of length $\Gamma = 20$ produced stable SOC states. Their wavelength was always $\lambda = 2$. (a) Mixing parameter $M$, (b) Nusselt number. The SOC branch in a pure fluid is shown as the dotted curve. Thresholds are $r_{stat}(k_c = 0) = 0.04216$ and $r_{stat}(k = \pi) = 0.06012$ [68].

FIG. 12. Snapshots of the first four SOC states of Fig. 11. The states at (a) $r = 0.506$ and (b) $r = 0.9116$ lie in the Soret region. Those at (c) $r = 1.1142$ and (d) $r = 2.026$ lie in the Rayleigh region. The lateral profiles at $z = 0.5$ show $w$ (thin line), $40\,\delta T$ (triangles), and $1000\,\delta C$ (squares). Otherwise the presentation is as in Fig. 3.

FIG. 13. Bifurcation diagrams of frequency and Nusselt number for $\psi = -0.25$ and four combinations of $L$ and $\sigma$ over a common interval of Rayleigh numbers. See also Tab. II. Stable TW's are presented by dots and SOC states by squares. Curves are guides to the eye. Unstable TW branches (dashed lines) connecting to $\omega_H$ in the upper row and to $r_{osc}$ (crosses) in the lower row are schematic. SOC states in the pure fluid, $\psi = 0$, are shown by dotted lines in the lower row.

FIG. 14. (a) Frequency of nonlinear states at $r_{osc}$ on the upper bifurcation branch versus Lewis number. Note the logarithmic scale of $L$. (b) $L$-$r$ phase diagram. Symbols denote final states obtained in simulations. The dashed curves $r^*$ and $r_{TW}^s$ denoting the borders of the dashed area of stable TW's are hand-drawn interpolations. The thin curve, $r_{osc}$ [68], is the TW bifurcation threshold. Some properties of the phase diagram are given in Tab. III.

FIG. 15. (a) Frequency of nonlinear states on the upper bifurcation branch at $r_{osc}$ versus Prandtl number. Note the logarithmic scale of $\sigma$. (b) $\sigma - r$ phase diagram. Symbols represent final states obtained in simulations. The dashed curves, $r^*$ and $r_{TW}^s$, denoting the borders of the dashed area of stable TW's are hand-drawn interpolations. The thin curve, $r_{osc}$ [68], is the TW bifurcation threshold. Some properties of the phase diagram are given in Tab. IV.

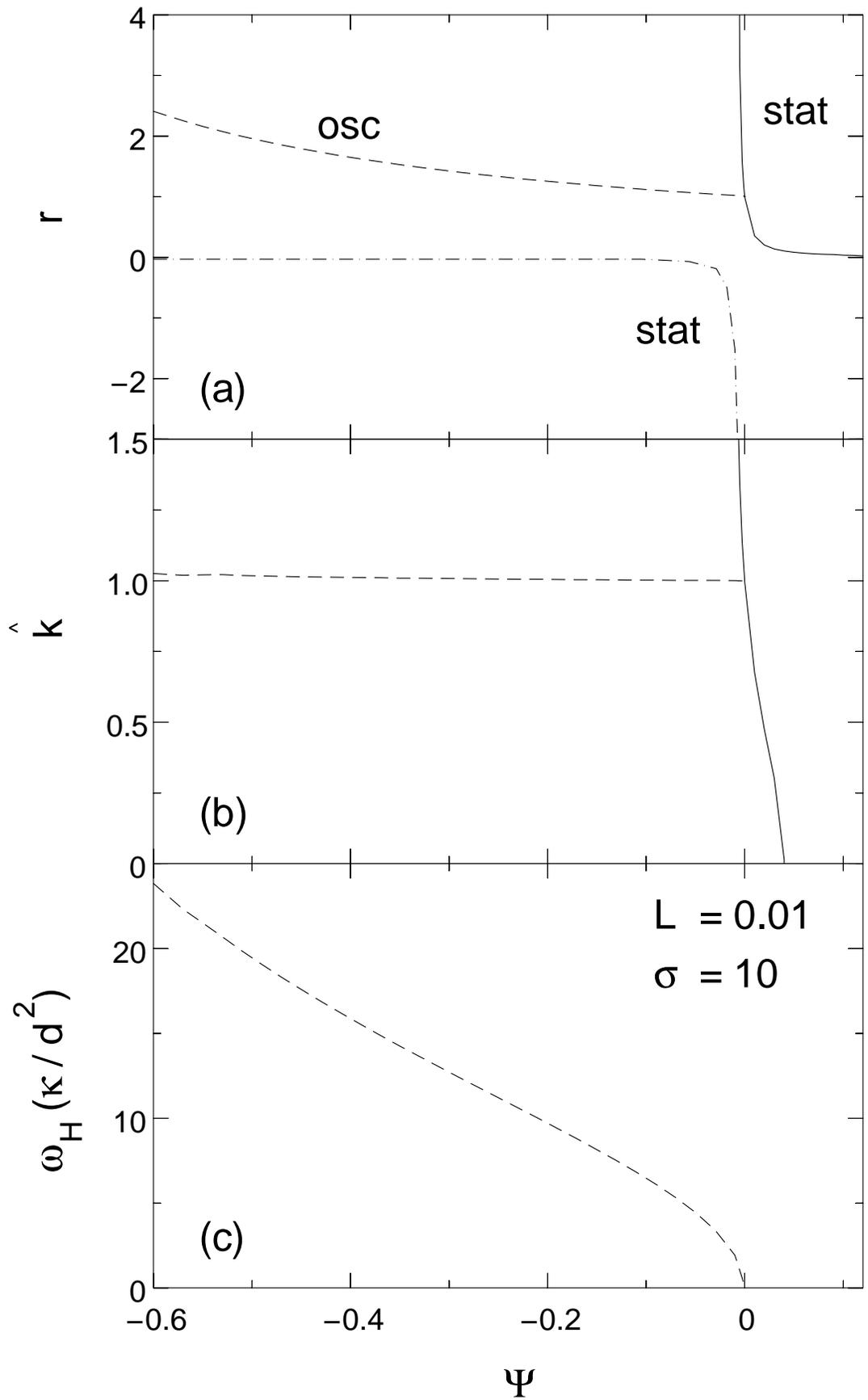

Fig. 1
Barten et al., Convection in Binary Fluid Mixtures I, Phys. Rev. E

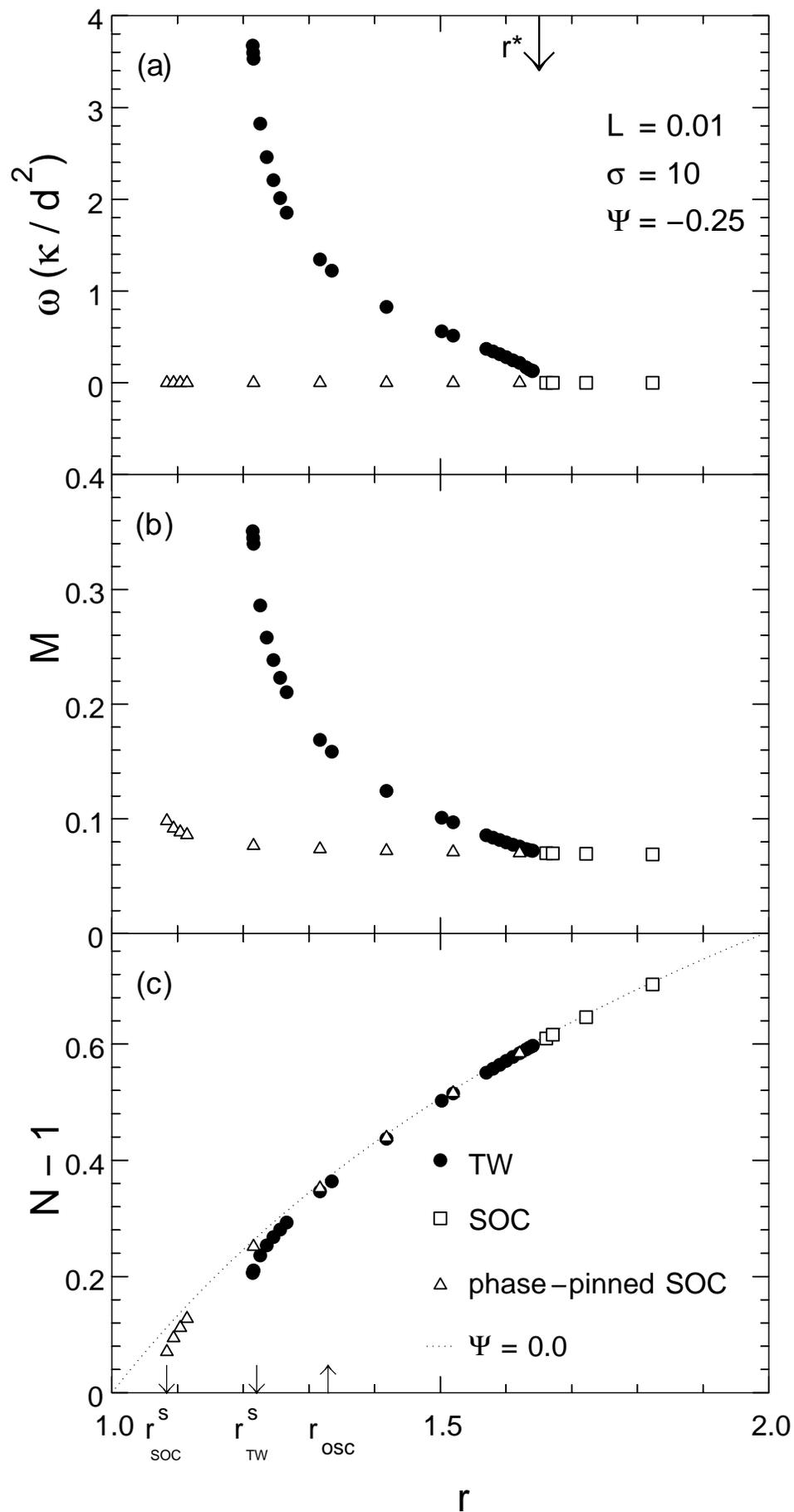

Fig. 2
Barten et al., Convection in Binary Fluid Mixtures I, Phys. Rev. E

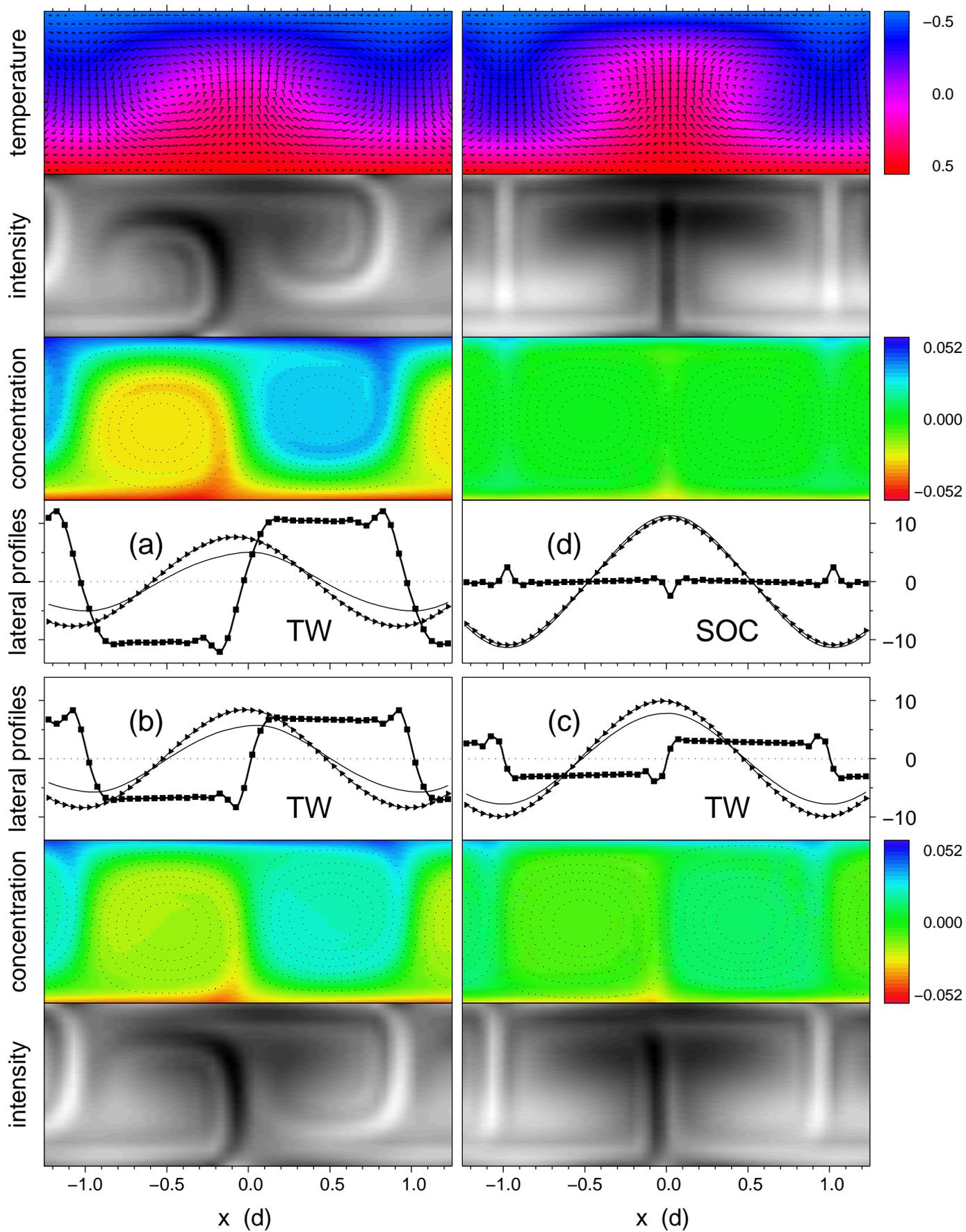

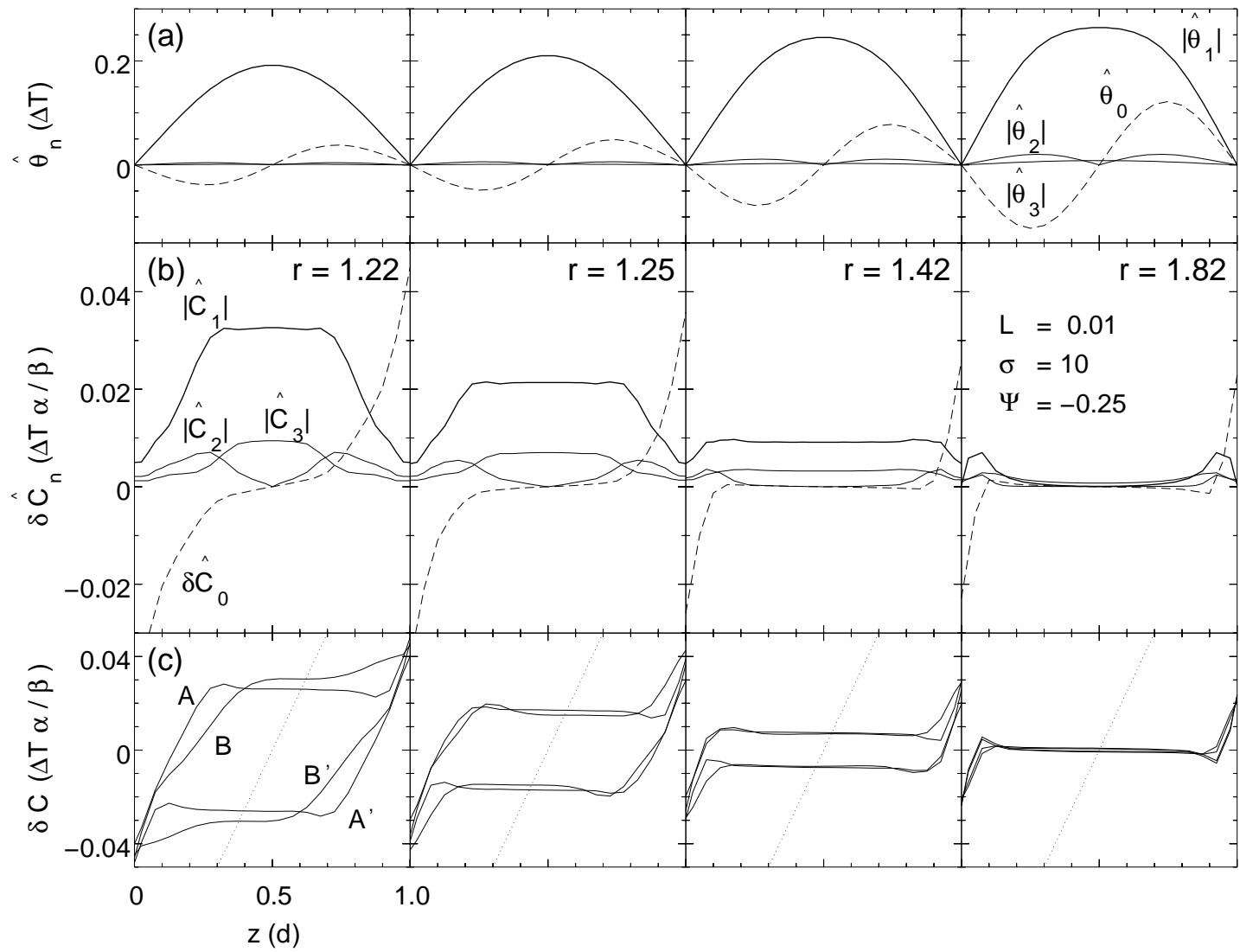

Fig. 4
Barten et al., Convection in Binary Fluid Mixtures I, Phys. Rev. E

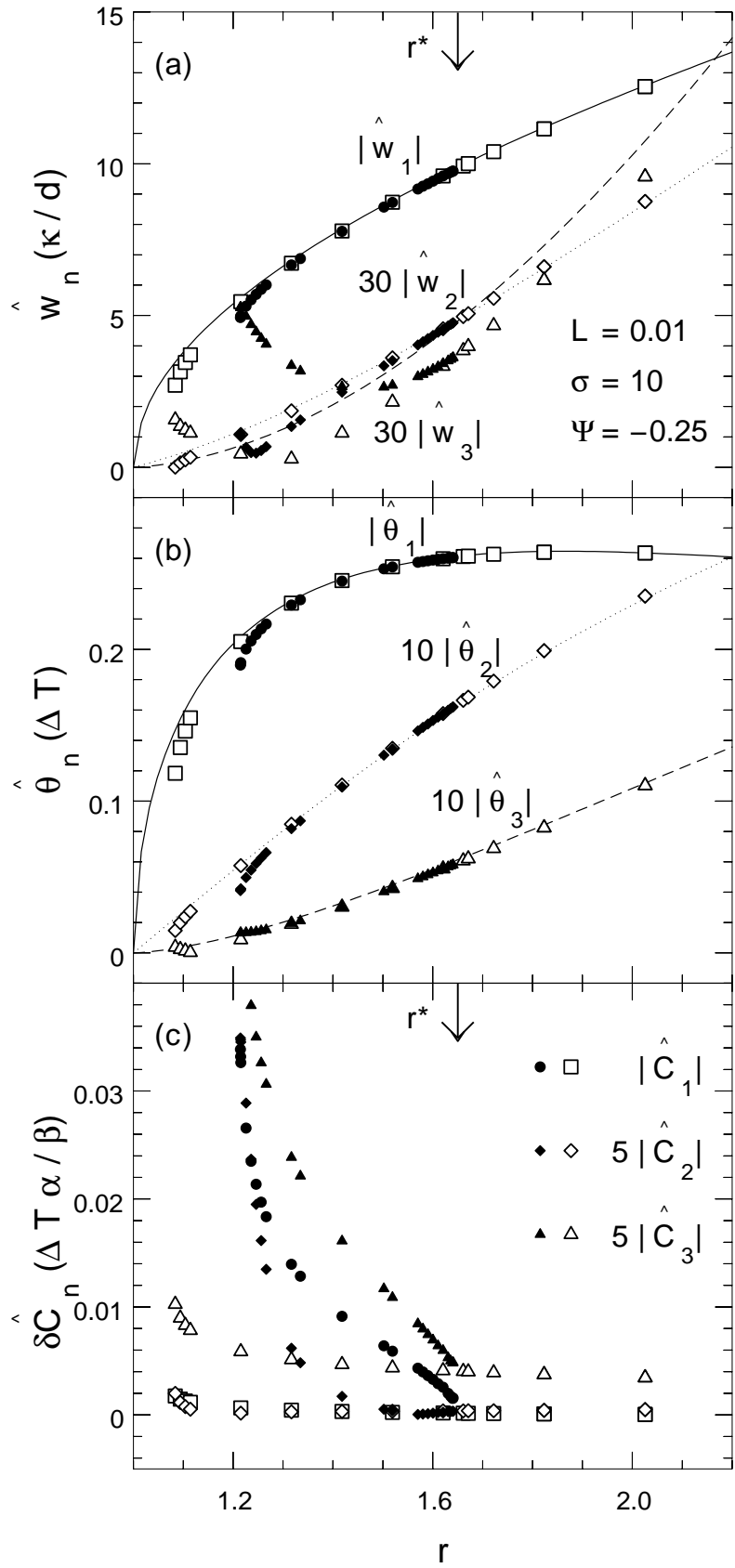

**Fig. 5**
Barten et al., Convection in Binary Fluid Mixtures I, Phys. Rev. E

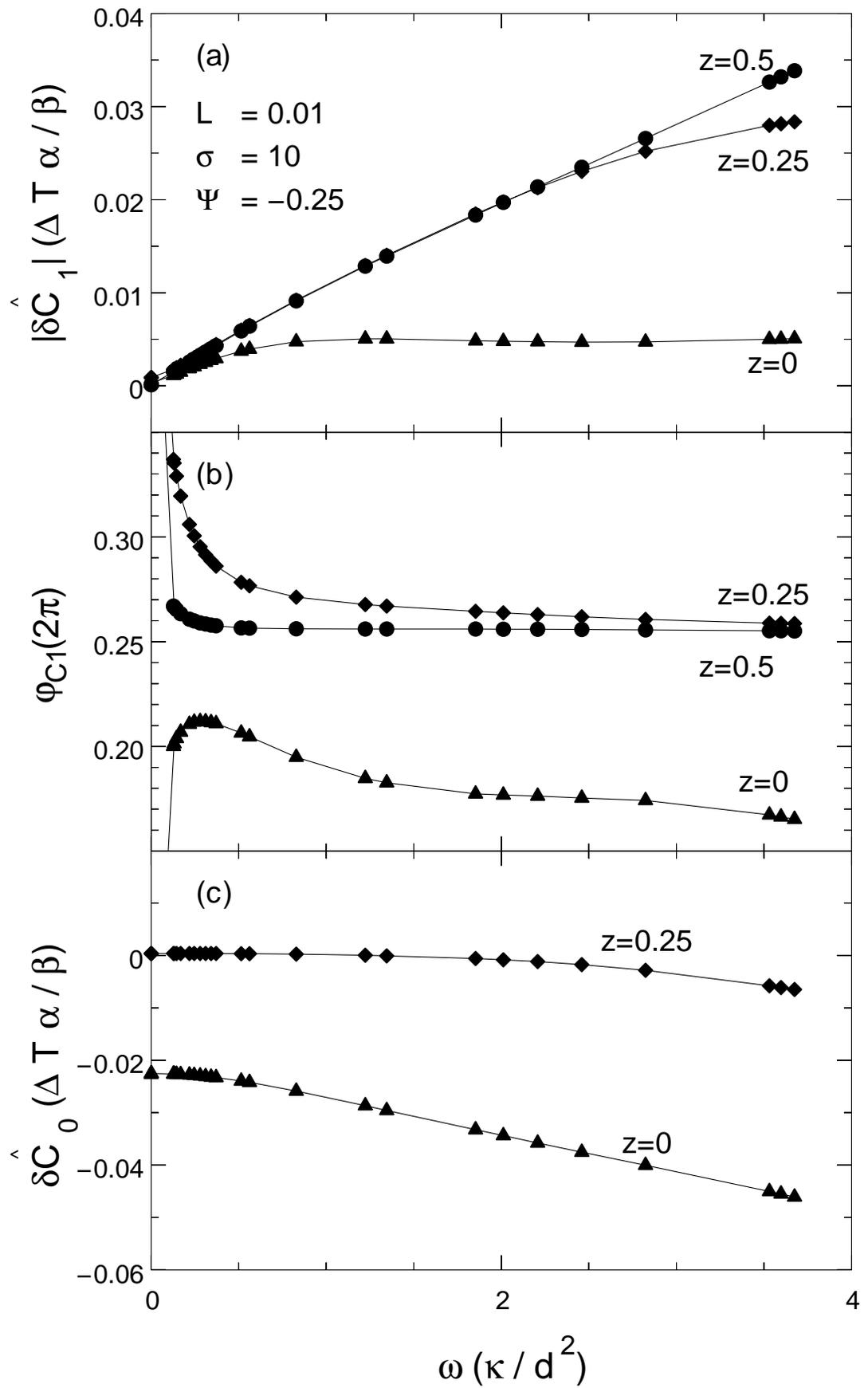

Fig. 6
Barten et al., Convection in Binary Fluid Mixtures I, Phys. Rev. E

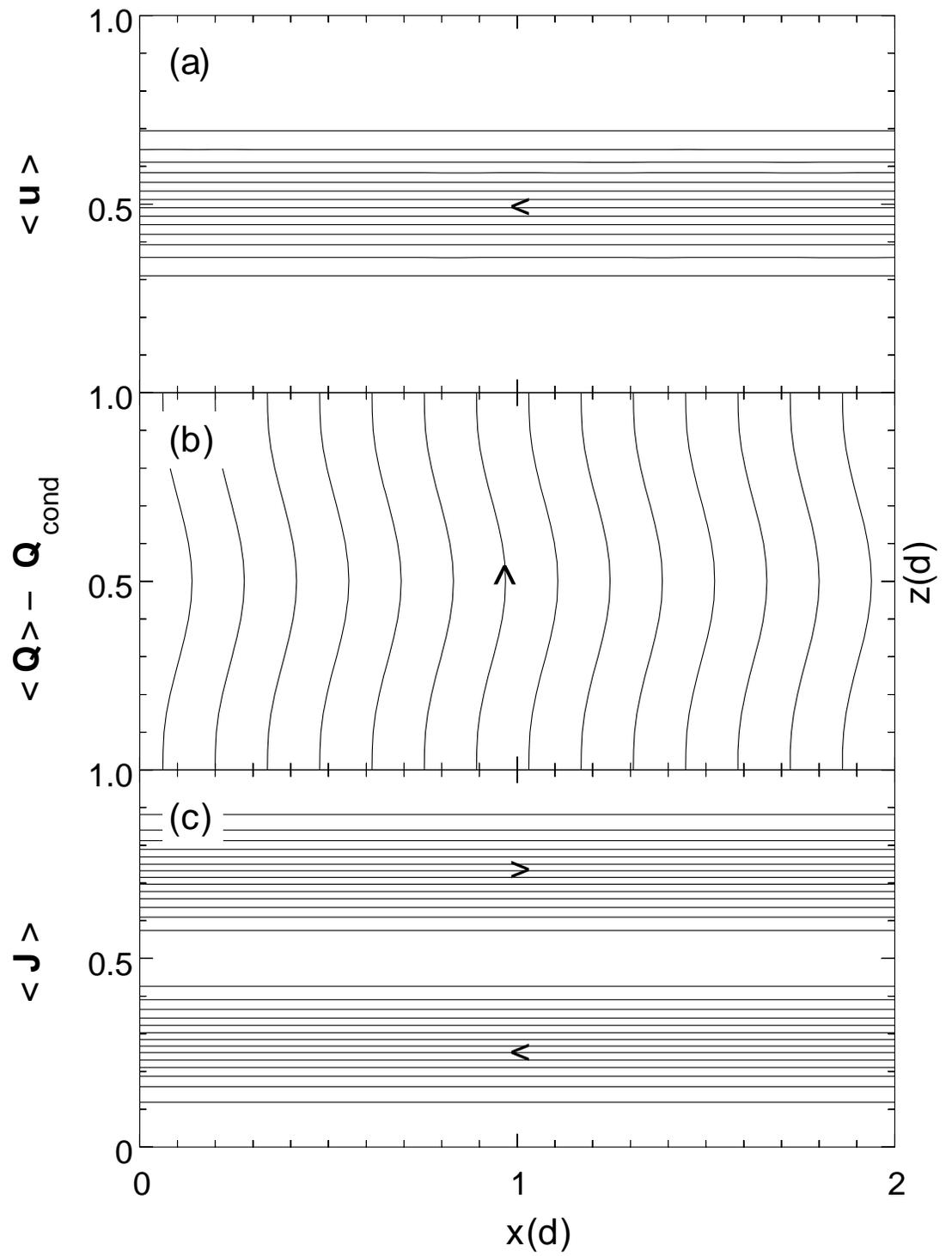

Fig. 7
Barten et al., Convection in Binary Fluid Mixtures I, Phys. Rev. E

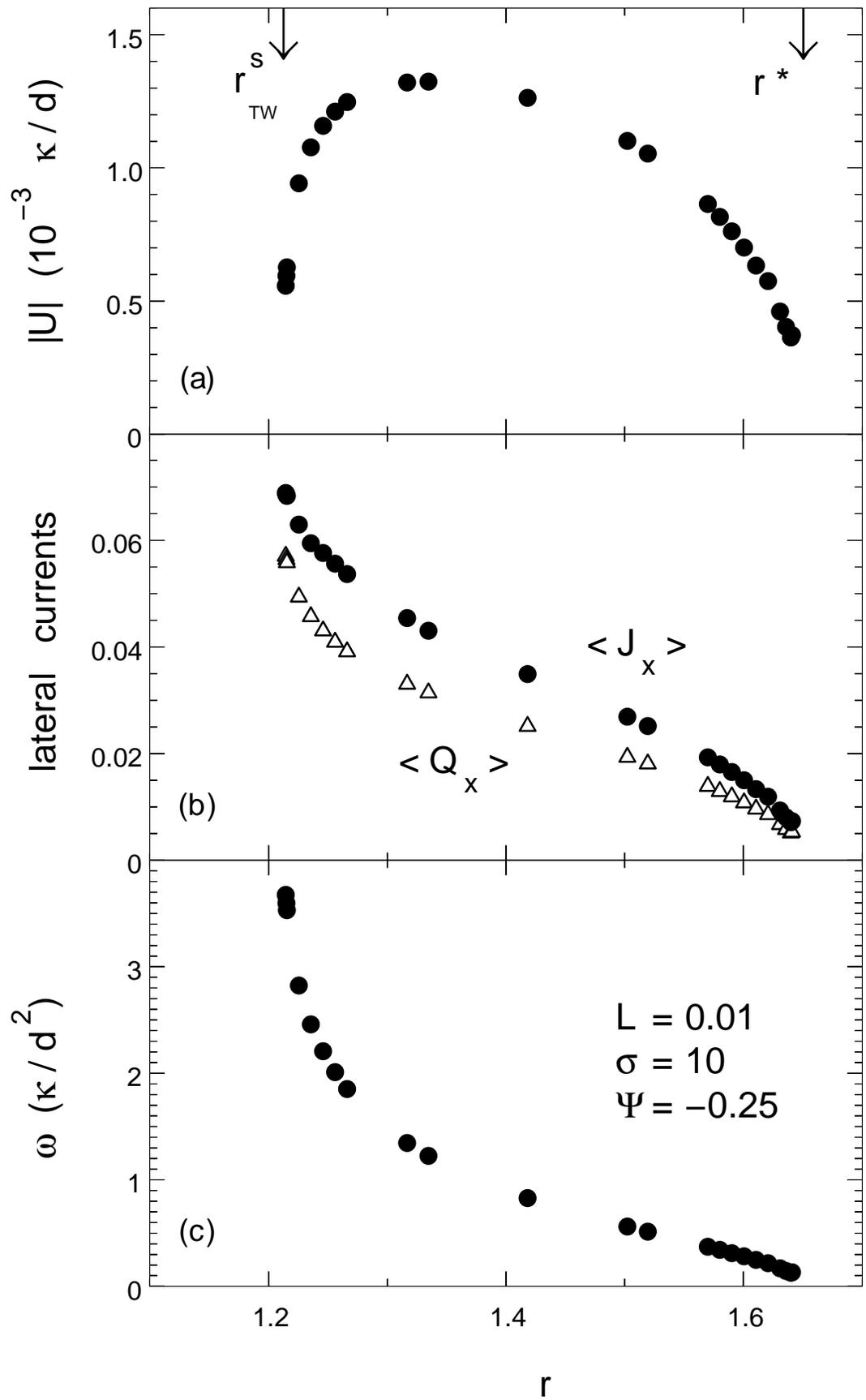

Fig. 8
Barten et al., Convection in Binary Fluid Mixtures I, Phys. Rev. E

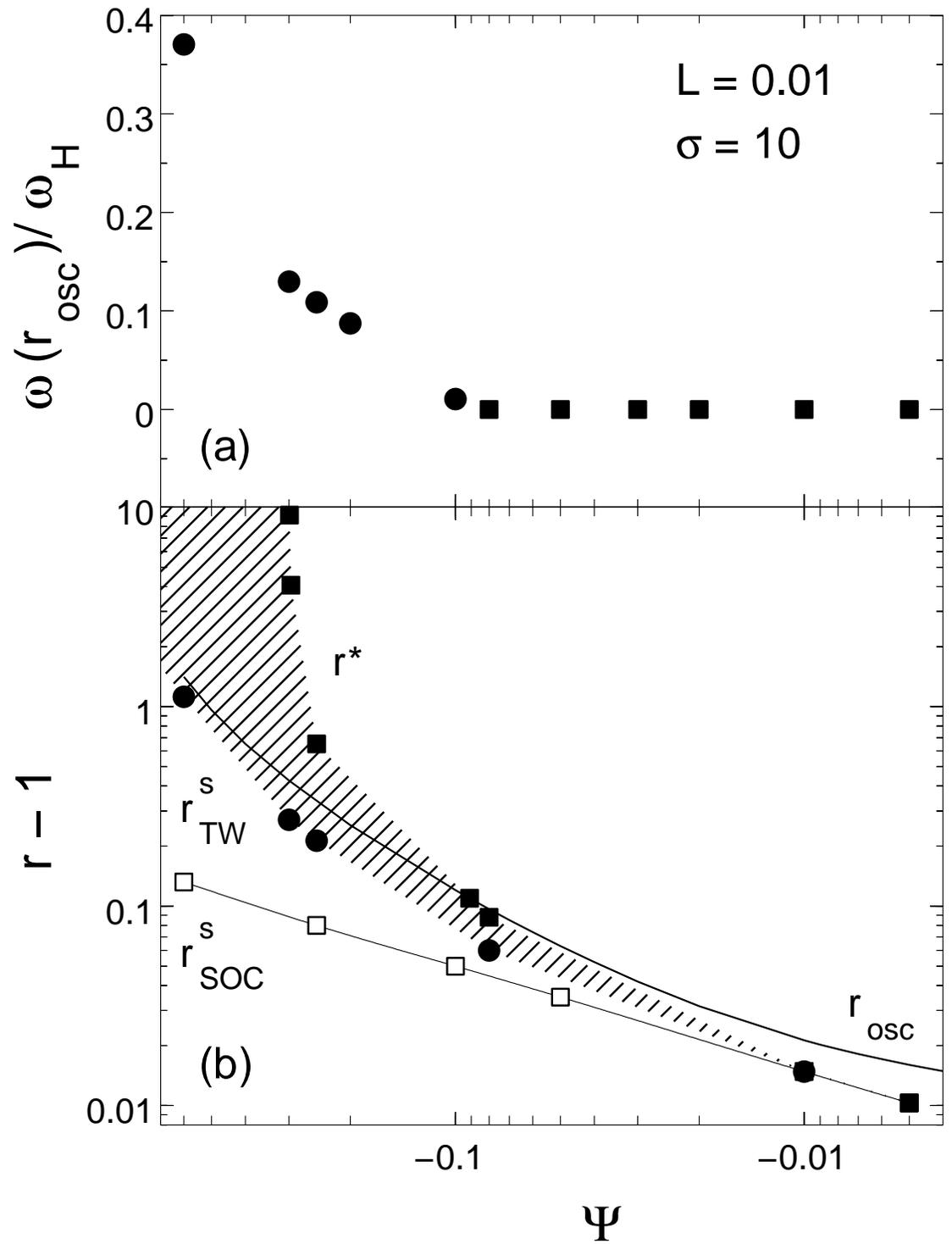

Fig. 9
Barten et al., Convection in Binary Fluid Mixtures I, Phys. Rev. E

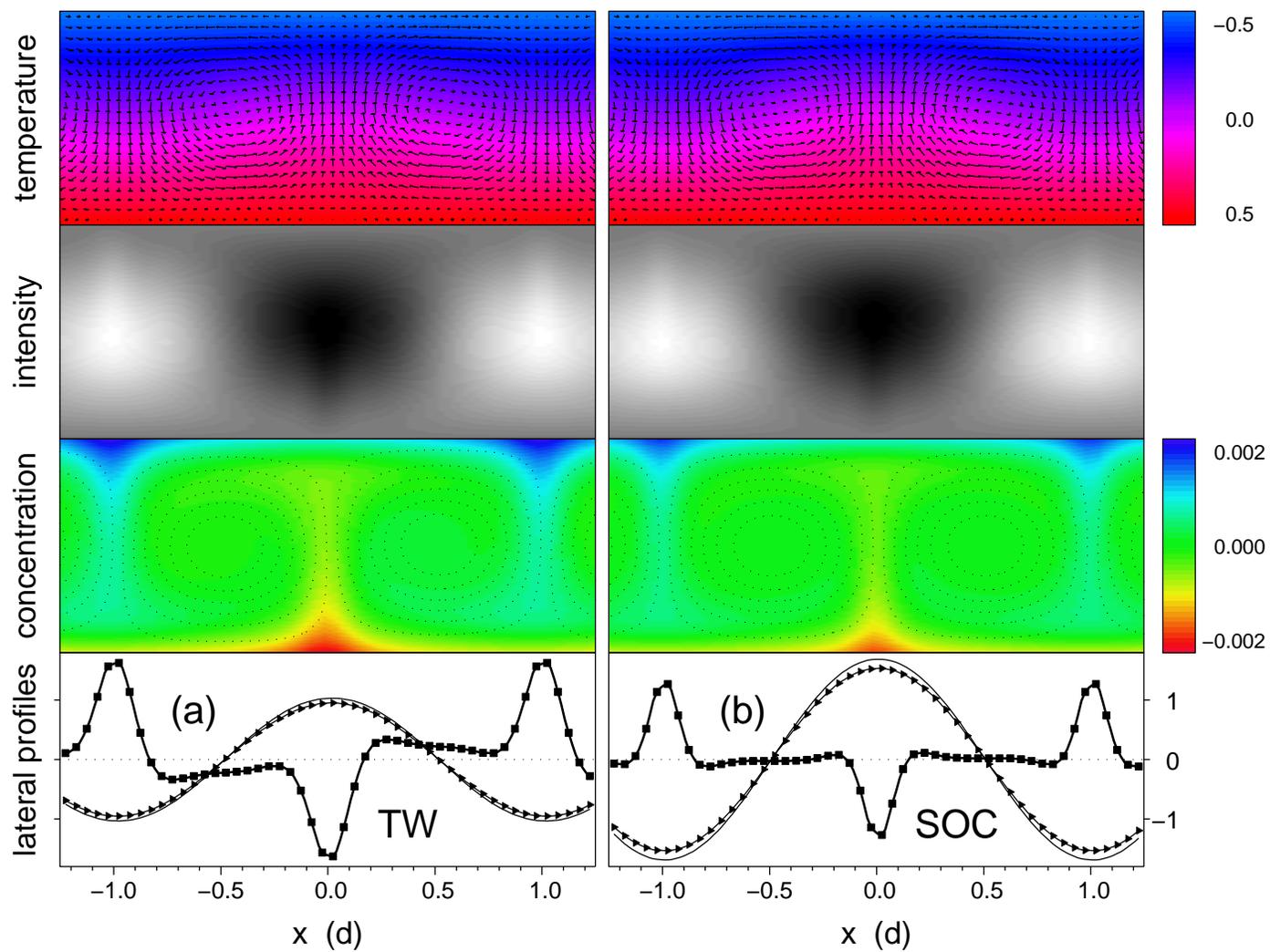

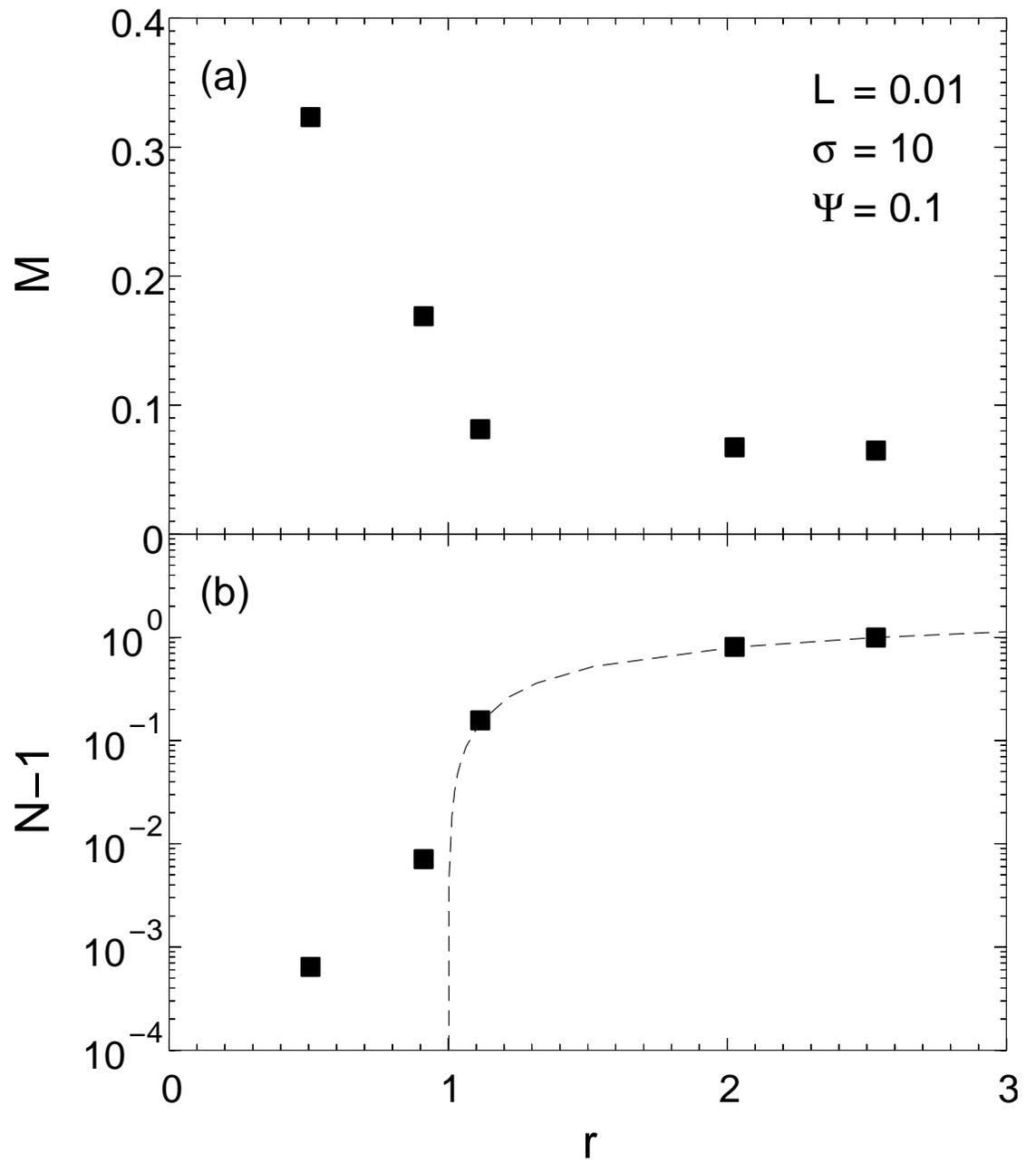

Fig. 11
Barten et al., Convection in Binary Fluid Mixtures I, Phys. Rev. E

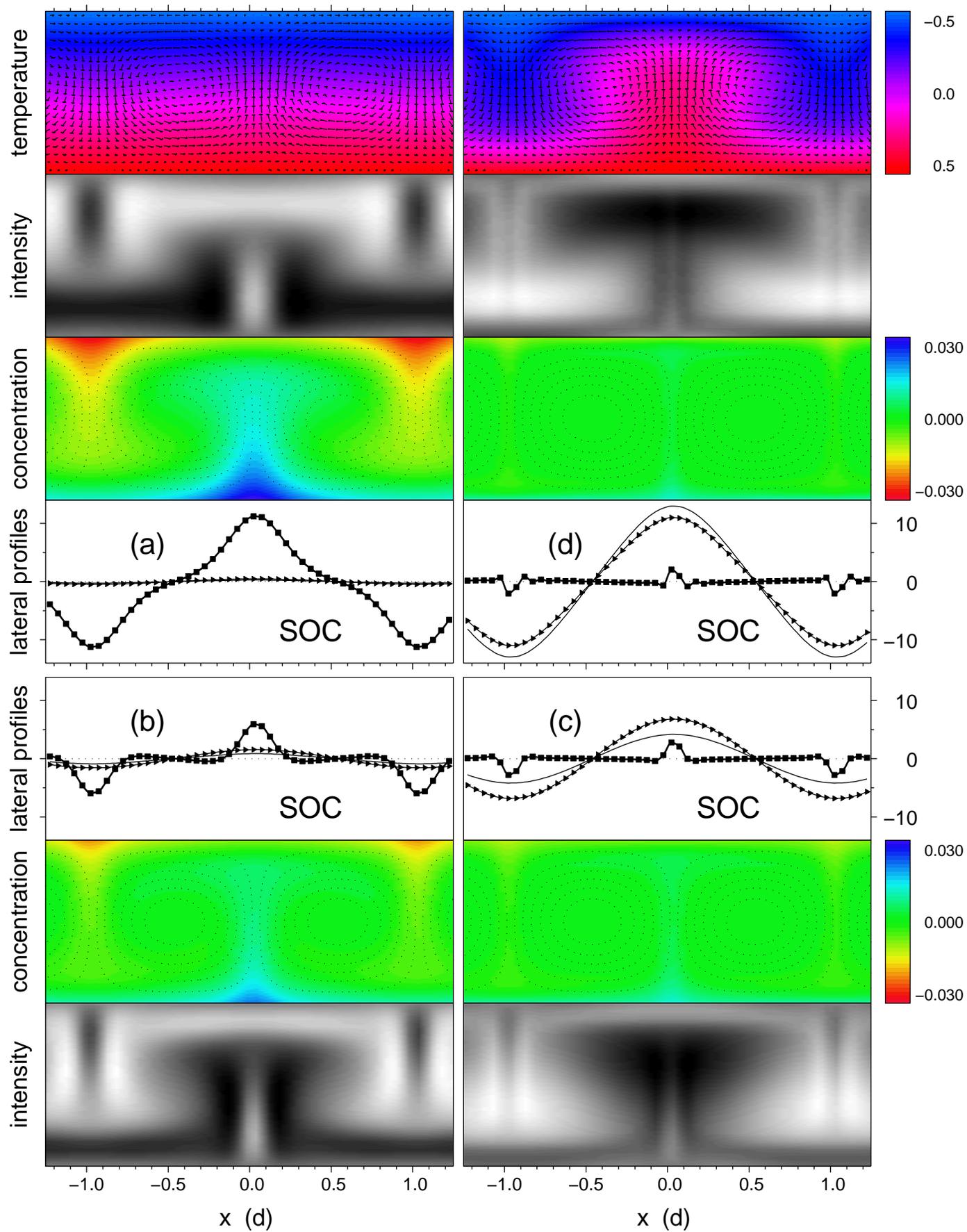

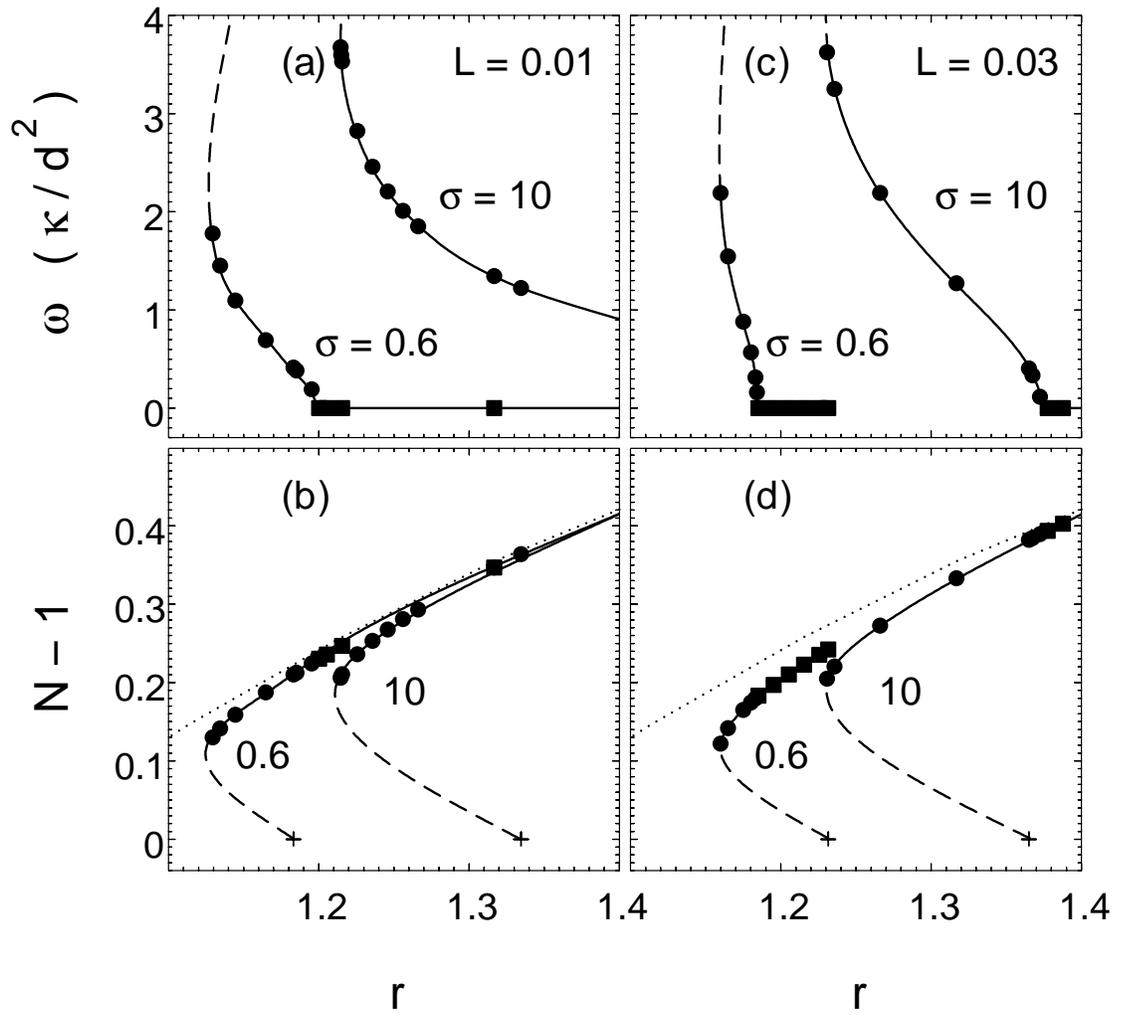

Fig. 13
Barten et al., Convection in Binary Fluid Mixtures I, Phys. Rev. E

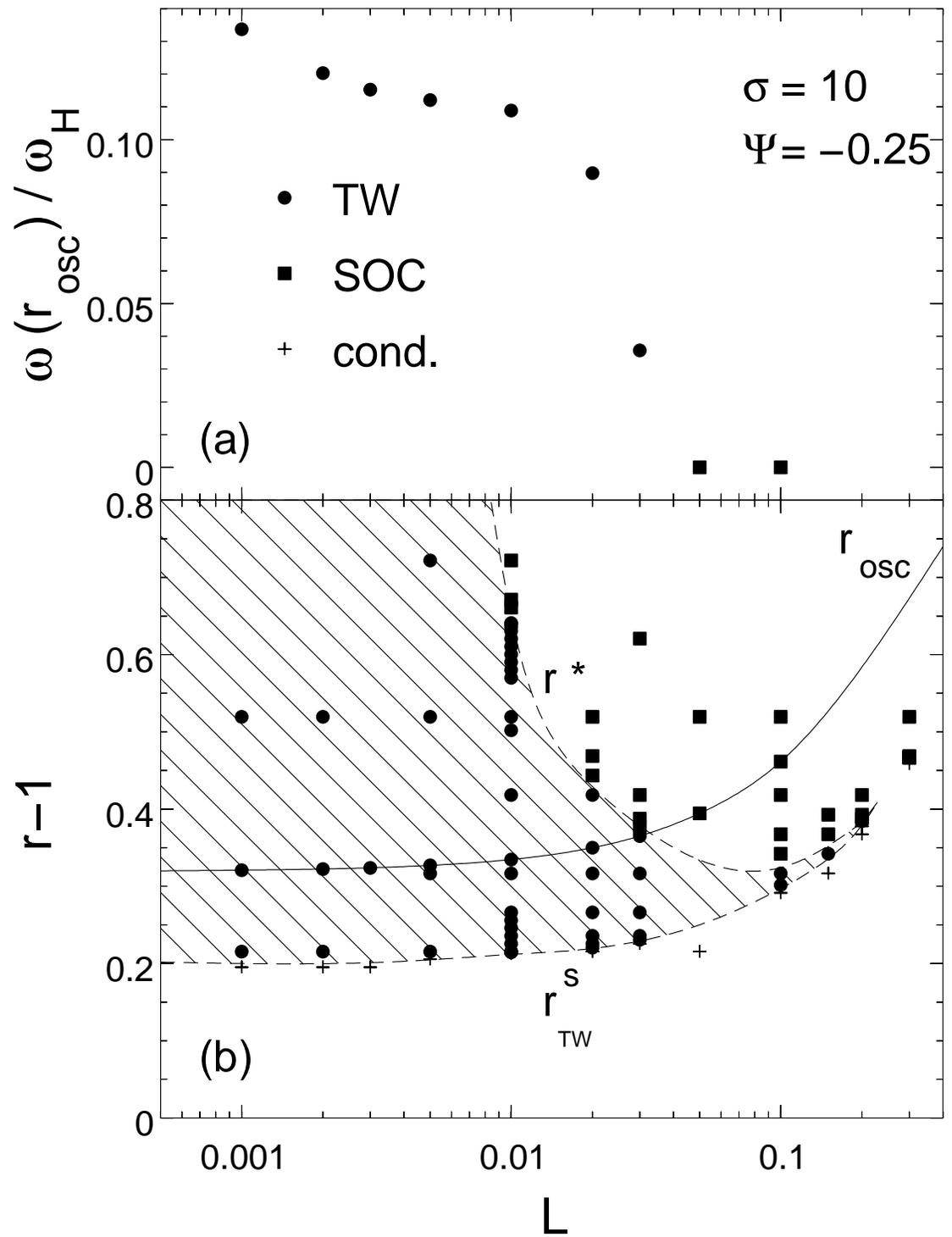

Fig. 14
Barten et al., Convection in Binary Fluid Mixtures I, Phys. Rev. E

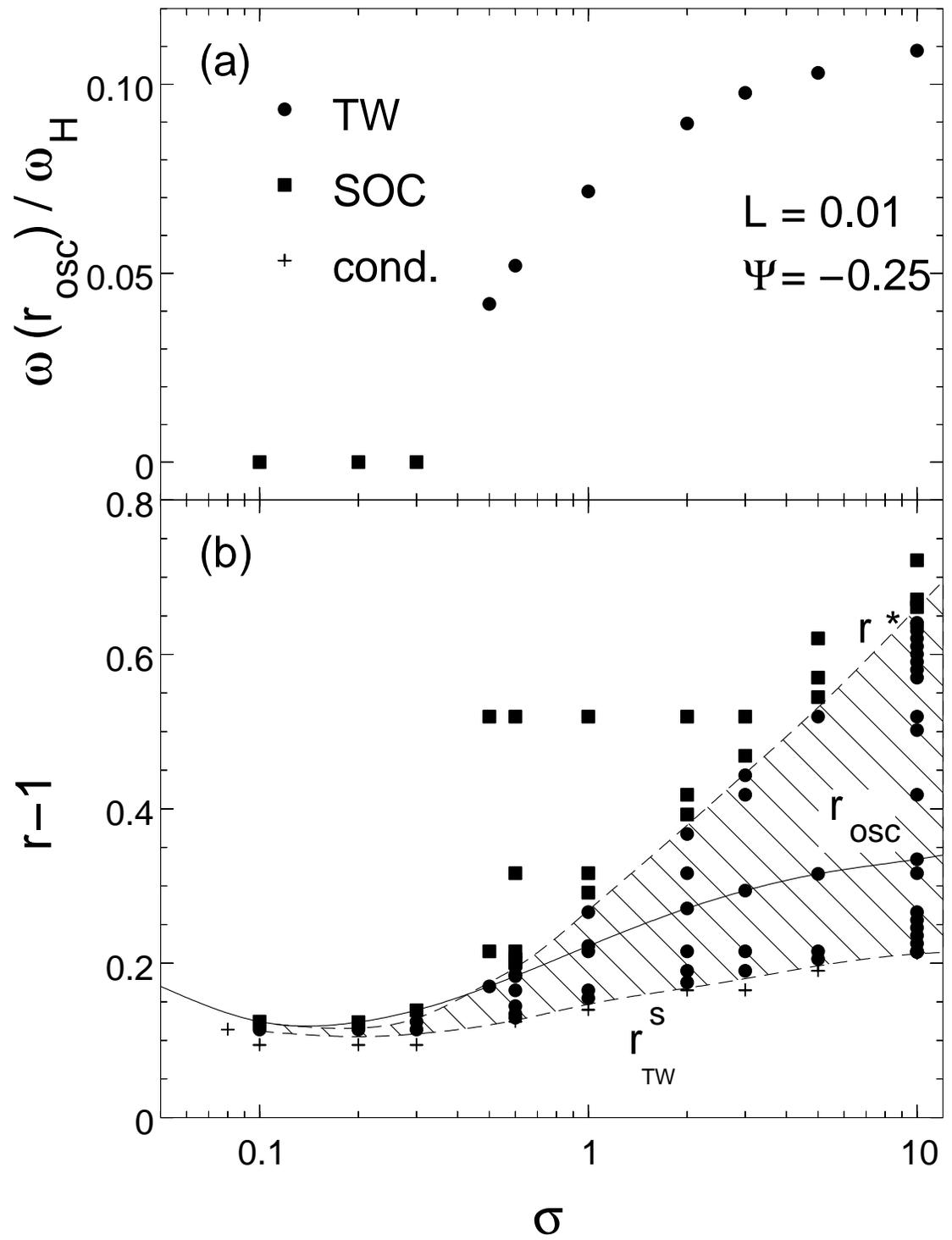

Fig. 15
Barten et al., Convection in Binary Fluid Mixtures I, Phys. Rev. E